\documentclass[a4paper,11pt]{article}

\usepackage{jheppub} 
\usepackage[T1]{fontenc} 
\usepackage[textsize=footnotesize,textwidth=2.5cm]{todo notes}

\makeatletter

\newcommand{\Rmnum}[1]{\expandafter\@slowromancap\romannumeral #1@}
\makeatother
\definecolor{mikadoyellow}{rgb} {0.16, 0.254, 0.6}
\usepackage{amssymb,amsmath,latexsym,bm,amsfonts}
\usepackage{graphicx}
\usepackage{longtable}
\usepackage{color,xcolor}
\usepackage{float}
\usepackage{indentfirst}
\usepackage{subfigure}
\usepackage{comment}
\usepackage{verbatim}
\usepackage{tikz}
\usepackage{colortbl}
\definecolor{LouisBlue}{RGB}{55, 114, 202}
\definecolor{LouisOrange}{RGB}{180, 54, 22}
\definecolor{LouisColor1}{RGB}{0, 118, 63}
\definecolor{LouisColor2}{RGB}{111, 73, 189}
\definecolor{lightblue}{RGB}{135, 206, 250}
\definecolor{yellowr}{RGB}{250, 250, 210}

\newcommand{\be}{\begin{equation}}
\newcommand{\ee}{\end{equation}}
\newcommand{\bpm}{\begin{pmatrix}}
\newcommand{\epm}{\end{pmatrix}}

\newcommand{\abs}[1]{|#1|}

\newcommand{\beqn}{\begin{eqnarray}}
\newcommand{\eeqn}{\end{eqnarray}}

\newcommand{\besub}{\begin{subequations}}
\newcommand{\eesub}{\end{subequations}}
\newcommand{\bea}{\begin{eqnarray}}
\newcommand{\eea}{\end{eqnarray}}

\def\is{\mathrm{Is}}

\definecolor{LouisBlue}{RGB}{55, 114, 202}
\definecolor{LouisOrange}{RGB}{180, 54, 22}
\definecolor{LouisColor1}{RGB}{0, 118, 63}
\definecolor{LouisColor2}{RGB}{111, 73, 189}
\usetikzlibrary{matrix,shapes,fit,tikzmark,calc,3d,arrows.meta,scopes}
\usetikzlibrary{decorations.pathreplacing,decorations.markings,snakes,perspective,shadows,intersections}
\tikzset{middlearrow/.style={
		decoration={markings,
			mark= at position #1 with {\arrow{latex}} ,
		},
		postaction={decorate}
	}
}
\tikzset{inversemiddlearrow/.style={
		decoration={markings,
			mark= at position #1 with {-\arrow{latex[reversed]}} ,
		},
		postaction={decorate}
	}
}

\title{Partial entanglement entropy threads in island phase}

\author[a]{Qiang Wen,}
\author[a]{Mingshuai Xu,}
\author[a]{Haocheng Zhong}

\affiliation[a]{Shing-Tung Yau Center and School of Physics, Southeast University, Nanjing 210096, China}

\emailAdd{wenqiang@seu.edu.cn, xumingshuai@seu.edu.cn, zhonghaocheng@outlook.com}

\abstract{In the context of AdS/CFT, it was recently proposed that the boundary partial entanglement entropy structure can be represented by the so-called partial entanglement entropy (PEE) threads in the AdS bulk, which are bulk geodesics with the density determined by the boundary PEE structure \cite{Lin:2023rxc,Lin:2024dho}. In Poincar\'e AdS space, it was shown that the PEE threads cover the AdS space uniformly, such that the number of intersections between any bulk surface and the bulk PEE threads is always given by the area of the surface divided by 4G. 
	
In this paper, we investigate the configurations of PEE threads when the boundary state is in island phase. The island phase was studied in the context of the holographic Weyl transformed CFT$_2$, which has been shown to capture all the main features of AdS/BCFT. Compared with AdS$_3$/CFT$_2$, in island phase instead of modifying the distribution of the bulk PEE threads, we should replace the boundary points with the corresponding cutoff spheres. Then the two-point and four-point functions of twist operators can be reproduced by identifying the bulk homologous surfaces anchored on the corresponding cutoff spheres that has the minimal number of intersections with the bulk PEE threads. This gives us a better understanding about the PEE structure in island phase and reproduces the island formula for entanglement entropy by allowing homologous surfaces to anchor on any cutoff surfaces. Furthermore, it gives a demonstration for the two basic proposals and a better understanding for the entanglement contribution that makes the foundation to compute the balanced partial entanglement entropy (BPE) \cite{Basu:2023wmv} which reproduces the entanglement wedge cross-section in island phase.}

\begin{document} 
	\maketitle
	\flushbottom
	
	\section{Introduction}
    In AdS/CFT \cite{Maldacena:1997re,Gubser:1998bc,Witten:1998qj}, the Ryu-Takayanagi (RT) formula \cite{Ryu:2006bv,Ryu:2006ef,Hubeny:2007xt} is the starting point for us to understand the geometry in the dual gravity via the entanglement structure of the boundary CFT. The RT formula tells us that, the entanglement entropy $S_{A}$ for any region $A$ in the boundary CFT is captured by the area of the co-dimension two surface $\mathcal{E}_{A}$ in AdS bulk, which is homologous to $A$ and has the minimal area,
    \begin{align}
    S_{A}=\frac{Area(\mathcal{E}_{A})}{4G}\,.
    \end{align}
    The RT formula was later proved in \cite{Lewkowycz:2013nqa,Dong:2016hjy} via the replica trick applied to the dual gravity. It was further refined as the quantum extremal surface (QES) formula \cite{Engelhardt:2014gca} by including the quantum correction \cite{Faulkner:2013ana} for the holographic entanglement entropy.
    
    The RT formula interprets the area of bulk minimal surfaces anchored on the boundary as entanglement entropies in the dual field theory. It is interesting to ask to what extent can we reconstruct the bulk geometry from boundary information. Firstly, given the background geometry of the gravity, how can we interpret a generic geometric quantity, for example an arbitrary point, curve or surface in the bulk spacetime, in terms of quantum information quantities of the dual field theory? Secondly, how can we reconstruct the background geometry from the boundary information? So far, several approaches have been explored for this goal which are partially successful. See \cite{Chen:2021lnq} for a detailed review on the recent progresses on the program of geometry reconstruction in AdS/CFT.
     
    For the first question, given the geometry background one can interpret bulk curves via the so-called \textit{differential entropy} \cite{Balasubramanian:2013lsa,Headrick:2014eia,Czech:2014wka,Czech:2014ppa,Czech:2015qta,Czech:2015kbp}, by studying the set of geodesics tangent to the curve. Nevertheless, the reconstruction based on the \textit{differential entropy} turns out to be messy in higher dimensions \cite{Balasubramanian:2018uus}. Also in \cite{Freedman:2016zud,Headrick:2020gyq,Headrick:2022nbe} (see \cite{Agon:2021tia,Rolph:2021hgz,Caceres:2023ziv,Caggioli:2024uza,Du:2024xoz,Headrick:2020gyq,Headrick:2022nbe} for recent relevant research.), it was assumed that the boundary entanglement structure can be simulated by a set of \textit{bit threads} anchored on the boundary, which is described by a divergenceless vector field. This gives an equivalent formulation for the RT formula and interprets the area of the RT surface as the maximal flux of the \textit{bit threads} in AdS space out of that region. The \textit{bit threads} picture was also used to interpret the area of the (minimal) entanglement wedge cross section (EWCS) \cite{Bao:2019wcf}.
    
    For the second question, one can explore the bulk AdS metric by studying the toy models of quantum gravity based on \textit{tensor networks} \cite{Swingle:2009bg,evenbly2011tensor,Haegeman:2011uy,Swingle:2012wq,Qi:2013caa,Pastawski:2015qua,Hayden:2016cfa,Bhattacharyya:2016hbx,Bhattacharyya:2017aly}. The tensor networks are toy models that can reproduce some of the important features of AdS/CFT. For example, the entanglement entropy in the tensor network is captured by the homologous path in the network which cut the network with the minimal number of times, hence giving an analogue of the RT formula \cite{Swingle:2009bg,Hayden:2016cfa}. In \cite{Nozaki:2012zj} it was found that the bulk AdS geometry can emerge from pure CFT data using the continues multi-scale entanglement renormalization ansatz (cMERA) tensor network. Another way to get AdS metric is to optimize the path integral that computes the wave functionals in CFTs via Weyl transformations \cite{Caputa:2017urj}.

    Inspired by the above interesting concepts, the authors of \cite{Lin:2023rxc,Lin:2024dho} proposed a framework to reconstruct generic geometric quantities in AdS based on a new measure of entanglement, the \textit{partial entanglement entropy} (PEE) \cite{Wen:2018whg,Wen:2019iyq,Wen:2020ech,Han:2019scu,Han:2021ycp}\footnote{The relation between the PEE and the above mentioned concepts has been discussed in \cite{Kudler-Flam:2019oru,Wen:2018mev,Abt:2018ywl,Ageev:2019mbb,Rolph:2021nan,Gong:2023vuh,Lin:2022aqf,Lin:2021hqs,Lin:2023jah,Lin:2023hzs,Lin:2023orb}. Also, there are studies on reconstructing the EWCS using the PEE in various scenarios \cite{Wen:2021qgx,Camargo:2022mme,Wen:2022jxr,Basu:2022nyl,Basu:2023wmv,Lin:2023ajt}. }. The basic setup for this framework is that, we geometrize the two-point PEE between any two boundary points in terms of the bulk geodesic connecting these two points, which we call the \textit{PEE thread}. Hence, we get a network of PEE  threads consisting of all the bulk geodesics anchored on the boundary, and the density of the geodesics is determined by the PEE structure of the boundary theory. Unlike the \textit{bit threads}, the configuration of the PEE threads is totally determined by the boundary state and the PEE threads intersecting with each other. It is remarkable to find that for any bulk point, the number of the PEE threads passing through it is exactly $1/4G$. Then for any bulk surface, the number of intersections it intersect with the bulk PEE threads equals to the area of the surface multiplied by $1/4G$. Interestingly, the RT formula can be reproduced by minimizing the number of intersections between all possible homologous surfaces of a boundary region and the bulk PEE threads. In other words, the bulk PEE thread configuration gives a finer geometric description for the entanglement structure of the boundary holographic CFT. We will give more details for this framework in the next section.
    
    Recently, a new rule to compute the entanglement entropy for the Hawking radiation of black holes after the Page time was proposed, namely the \textit{island formula} \cite{Penington:2019npb,Almheiri:2019psf,Almheiri:2019hni,Almheiri:2019qdq,Penington:2019kki}. The result is consistent with the Page curve hence helps us better understand the puzzle of black hole information lost \cite{Hawking:1974sw}. It is very interesting and surprising that, the island formula claims that when computing the entanglement entropy $S_A$ of a region $A$ in gravitational systems, we should consider the possibility of including a region Is($A$) outside $A$ and compute $S_A$ via the following formula \cite{Penington:2019npb,Almheiri:2019psf,Almheiri:2019hni}:
	\begin{equation}\label{island formula 1}
	S_A=\min \text{ext}_{\text{Is}\left(A\right)}\left\{\frac{\text{Area}\left(\partial \text{Is}\left(A\right)\right)}{4G}+\tilde{S}_{\text{bulk}}\left(A\cup \text{Is}\left(A\right)\right)\right\}
	\end{equation}
	where $\tilde{S}_{\text{bulk}}\left(A\cup \text{Is}\left(A\right)\right)$ is the von
	Neumann entropy of region $A\cup\text{Is}\left(A\right)$ in the quantum field theory with fixed background curved spacetime and $\partial \text{Is}\left(A\right)$ is the boundary of Is$\left(A\right)$. More explicitly, we should consider all possible spatial region Is$\left(A\right)$ then take extremal value of the summation inside the bracket of \eqref{island formula 1}. If there are multiple saddle points, the minimal one should be chosen, and the region Is$\left(A\right)$ in this case is called the \emph{entanglement island} of region $A$. Given a state of a gravitational system, if there exists any region $A$ with non-trivial entanglement island $\text{Is}(A)$ according to \eqref{island formula 1}, then the state is in island phase. Following the entanglement wedge reconstruction program, it is believed that for gravitational systems in island phase, the state of the island region Is$\left(A\right)$ can be reconstructed\footnote{See \cite{Penington:2019kki} for exploration on explicit reconstruction channels.} from the information in $A$. In \cite{Basu:2022crn} this property was refined as the \textit{self-encoding} property, and the \textit{self-encoding} property in systems beyond gravitational theories is also discussed. 
	
	Inspired by the self-encoding property, it was proposed and tested that the doubly holographic nature \cite{Almheiri:2019hni,Geng:2020qvw} of the Karch-Randall braneworld \cite{Karch:2000gx,Karch:2000ct,Geng:2023qwm} and the later proposed AdS/BCFT \cite{Takayanagi:2011zk,Fujita:2011fp} correspondence  (see \cite{Deng:2022yll,Geng:2022slq,Geng:2024xpj,Miao:2022kve} for an incomplete list of recent developments in these configurations), which are extensively used context where entanglement islands are studied, can be perfectly simulated by a Weyl transformed holographic CFT$_2$ \cite{Basu:2022crn,Basu:2023wmv,Lin:2023ajt,Chandra:2024bkn} (see also  \cite{Akal:2021foz,Suzuki:2022xwv,Suzuki:2022yru} for similar proposals). In this paper, we will combine the Weyl transformed CFT$_2$ setup with the PEE thread configuration. More explicitly, we will analyze the island configurations and the entanglement structure for the Weyl transformed CFT$_2$ based on a modified PEE thread configuration. One of our main task is to reproduce the island formula \eqref{island formula 1} in KR braneworld or AdS/BCFT via optimizing the number of intersections between the PEE threads and the surfaces homologous to boundary regions in the gravity dual of the Weyl transformed CFT$_2$. Furthermore, based on the PEE thread configuration, we give a holographic demonstration for the two basic proposals in \cite{Basu:2023wmv}, which relates the two-point and four-point functions of twist operators to the partial entanglement entropy in the Weyl transformed CFT$_2$. 
	
	The paper is organized as follows. We give a brief introduction to the notions of partial entanglement entropy and the corresponding PEE threads in section \ref{sec: PEE}, and then we introduce the holographic Weyl transformed CFT in section \ref{sec: Weyl CFT}. In section \ref{sec.PEEinislandphase}, we give a proposal for the PEE thread configuration which describes the entanglement structure of the holographic Weyl transformed CFT$_2$. Based on the new setup, we calculate the PEE in island phase and demonstrate the two basic proposals in \cite{Basu:2023wmv}. In section \ref{sec:PEE island} we reproduce the island formula \eqref{island formula 1} by optimizing the number of intersections between the PEE threads and homologous surfaces. We revisited the entanglement contribution and BPE in island phase and clarify the bases for the computations of the BPE in \cite{Basu:2023wmv} in section \ref{sec.revisited}. We summarize our results and give a discussion in the last section.

	\section{A brief introduction to PEE and PEE threads}\label{sec: PEE}
	
	\subsection{Partial entanglement entropy}
	
	The concept of PEE originates from the \textit{entanglement contour} function $s_A\left(x\right)$ \cite{Chen:2014}, which was conjectured to describe the contribution from each site $x$ inside a region $A$ to the entanglement entropy $S_A$. In this paper, we mainly focus on two-dimensional systems and $x$ characterizes the spatial direction. By definition, the entanglement entropy can be recovered by collecting the contributions from all sites inside region $A$:
	\begin{equation}
		S_A=\int_{A}s_{A}\left(x\right)dx.
	\end{equation}
	Note that the contour function is non-local due to the dependence on the region $A$. Then the contribution from a subset $\alpha\subset A$ to $S_A$ is calculated by,
	\begin{equation}
		s_{A}\left(\alpha\right)=\int_{\alpha}s_A\left(x\right)dx.
	\end{equation}
	Provided $\bar{A}\cup A$ makes a pure state, the \emph{partial entanglement entropy} (PEE)  $\mathcal{I}\left(\alpha,\bar{A}\right)$ \cite{Kudler-Flam:2019oru,Wen:2018whg,Han:2019scu,Wen:2019iyq,Wen:2020ech,Han:2021ycp} is a measure of two body correlation between $\alpha\subset A$ and the complement $\bar{A}$, which can be calculated by
	\begin{equation}\label{tworep}
		s_A\left(\alpha\right)\equiv\mathcal{I}\left(\alpha,\bar{A}\right),
	\end{equation}    
	and the entanglement contour function is a differential version of the PEE, see \cite{Basu:2023wmv,Wen:2021qgx,Ageev:2021ipd,Rolph:2021nan,Camargo:2022mme,Wen:2022jxr,Lin:2022aqf,Lin:2023orb} for recent progresses on PEE. We call the left-hand side the \textit{contribution representation} of the PEE which emphasizes the contribution from the subset $\alpha$, while we call the right-hand side the \textit{two-body-correlation}  representation of the PEE which emphasizes the two-body correlation \cite{Wen:2019iyq}. 
	
	Although the explicit definition of the PEE based on reduced density matrix is still missing, it must satisfy a set of requirements \cite{Chen:2014, Wen:2019iyq} including all the requirements satisfied by the mutual information $I\left(A, B\right)$\footnote{Note that, we should be careful not to confuse the PEE $\mathcal{I}\left(A,B\right)$ and the mutual information $I\left(A,B\right)$.} and the additional
	key requirement of additivity according to its interpretation based on the contour function. More explicitly, assuming $A,B,C$ are non-overlapping regions on a Cauchy surface, the requirements are given by, 
	\begin{itemize}
		\item[1.] \textit{Additivity}: $\mathcal{I}\left(A,B\cup C\right)=\mathcal{I}\left(A,B\right)+\mathcal{I}\left(A,C\right)$;
		\item[2.] \textit{Permutation symmetry}: $\mathcal{I}\left(A,B \right)=\mathcal{I}\left(B,A\right)$;
		\item[3.] \textit{Normalization}\footnote{Note that the requirement is subtle since the entanglement entropy is divergent in quantum field theory, hence we should choose an appropriate regularization scheme, see \cite{Han:2019scu} for early discussions. See also \cite{SinghaRoy:2019urc} for a relevant discussion.}: $\mathcal{I}\left(A,\bar{A}\right)=S_A$;
		\item[4.] \textit{Positivity}: $\mathcal{I}\left(A,B\right)>0$;
		\item[5.] \textit{Upper bounded}: $\mathcal{I}\left(A,B \right)\leq \min\left\{S_A,S_B\right\}$;
		\item[6.]\textit{$\mathcal{I}\left(A,B\right)$ should be invariant under local unitary transformations inside $A$ or $B$};
		\item[7.] \textit{Symmetry: For any symmetry transformation $\mathcal{T}$ under which $\mathcal{T}A=A'$ and $\mathcal{T}B=B'$, we have $\mathcal{I}\left(A,B\right)=\mathcal{I}\left(A',B'\right)$}.
	\end{itemize}
  Based on \cite{Casini:2008wt}, it was demonstrated in \cite{Wen:2019iyq} that the above requirements have unique solution for Poincar\'e invariant theories. More interestingly, when we only apply the property of normalization to spherical regions \cite{Lin:2023rxc,Lin:2024dho}, the explicit formula of the solution can be determined by these requirements for the vacuum state of conformal field theories on a plain. Due to additivity and permutation symmetry of the PEE, any PEE $\mathcal{I}\left(A,B\right)$ can be
  evaluated as the integration of the \emph{two-point PEE} $\mathcal{I}\left(x,y\right)$ \cite{Wen:2019iyq}:
  \begin{equation}\label{two-point PEE 2}
  	\mathcal{I}\left(A,B\right)=\int_{A}dx\int_{B}dy\:\mathcal{I}\left(x,y\right).
  \end{equation}
  Note that the two-point PEE is an intrinsic quantity describing the entanglement structure of the system, hence it is independent of the choice of the regions $A$ and $B$. For the vacuum CFT$_2$, the two-point PEE is given by \cite{Lin:2023rxc}
  \begin{equation}\label{two-point PEE}
  	\mathcal{I}\left(x,y\right)=\frac{c}{6}\frac{1}{\left(x-y\right)^2}.
  \end{equation}
  where $c$ is the central charge.
 
 There are also other prescriptions to construct the PEE which satisfies the above seven requirements, e.g. the Gaussian formula applies for Gaussian states of many-body system \cite{Chen:2014}, and the geometric construction applied to the holographic theories with a local modular Hamiltonian \cite{Wen:2018whg,Wen:2018mev}. In two dimensional systems, the most powerful construction is the additive linear combination (ALC) proposal \cite{Wen:2018whg,Wen:2019iyq,Kudler-Flam:2019oru} applied to generic systems with all degrees of freedom settled in a unique order (for example a line or a circle). More explicitly, consider an interval $A$ which is partitioned into three sub-intervals $A=\alpha_L\cup \alpha \cup \alpha_R$ where $\alpha$ is the middle one and $\alpha_L(\alpha_R)$ denotes the left(right) sub-interval of $\alpha$, the proposal claims that
	\begin{equation}\label{ALC}
		\text{ALC proposal: } s_A\left(\alpha\right)\equiv\mathcal{I}\left(\alpha,\bar{A}\right)=\frac{1}{2}\left(S_{\alpha_L\cup \alpha}+S_{\alpha\cup \alpha_R}-S_{\alpha_L}-S_{\alpha_R}\right)
	\end{equation}
The PEEs calculated by different proposals are highly consistent with each other \cite{Wen:2018whg,Wen:2020ech,Kudler-Flam:2019nhr}.

	\subsection{PEE threads, PEE network and RT formula reformulation}\label{sec:PEE threads, PEE network}
\begin{figure}
		\centering
		\includegraphics[scale=0.25]{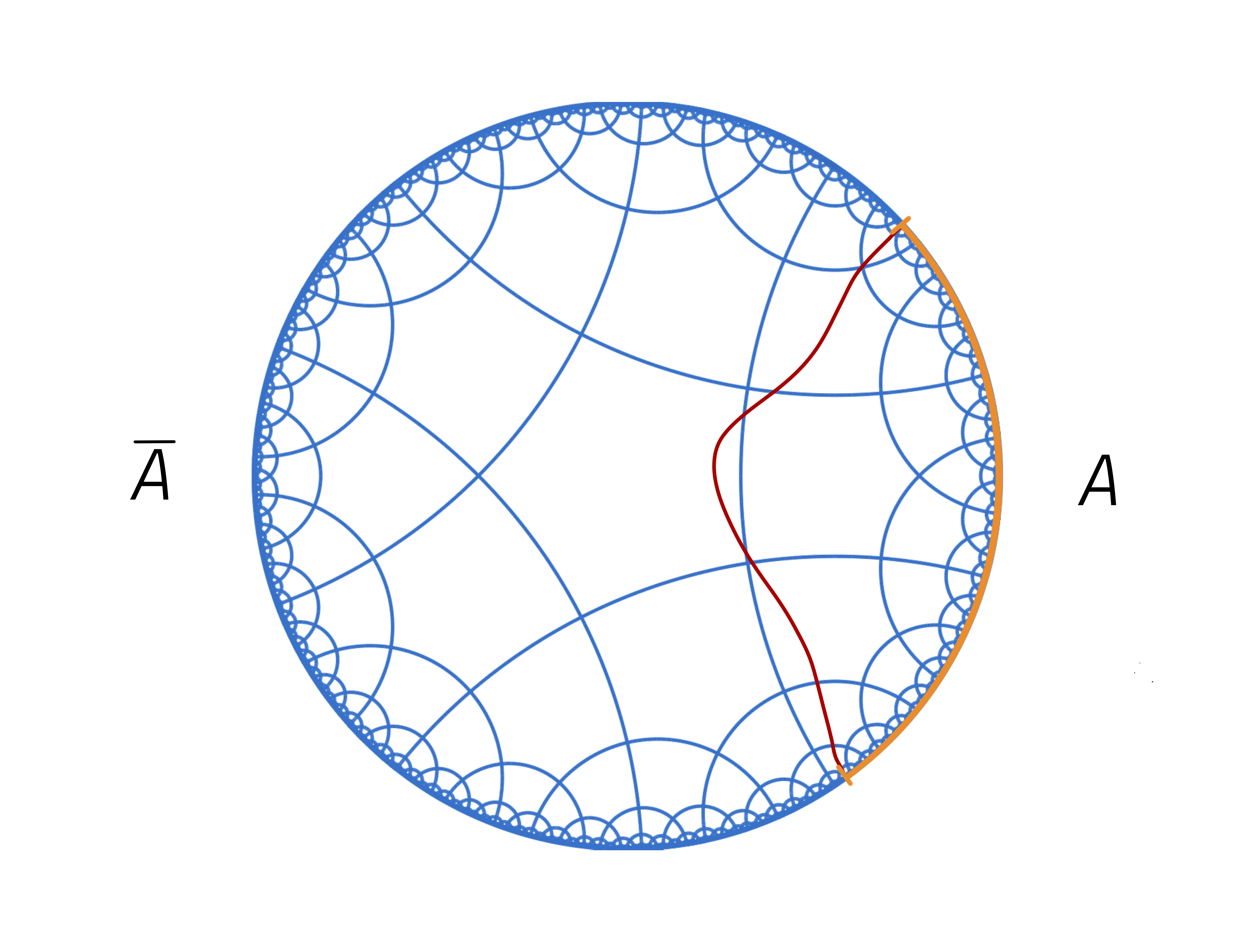}
		\caption{Visualization of PEE network on a time slice of global AdS$_3$ with boundary $A\cup \bar{A}$. For any arbitrary bulk surface homologous to $A$, one can count the number of intersections between the homologous surface (red line) and the PEE threads (blue lines).}
		\label{fig:PEE network}
	\end{figure}

	Recently, a scheme to geometrize the two-point PEE in the context of AdS/CFT was proposed in \cite{Lin:2023rxc,Lin:2024dho}. More explicitly, for the vacuum state of a holographic CFT$_d$ on a time slice, any two-point PEE $\mathcal{I}(\bf{x},\bf y)$ in the boundary CFT can be geometrized as a bulk geodesic connecting the two boundary points $\{\bf x, \bf y\}$. These bulk geodesics are called the \textit{PEE threads}. The set of all the PEE threads forms a continuous network in the bulk, which we call the \emph{PEE network} \cite{Lin:2023rxc,Lin:2024dho} (see Fig.\ref{fig:PEE network} for an example). The configuration of PEE threads in the AdS space emanating from any boundary point $\bf x$ can be described by a divergenceless vector field $V^\mu_{\bf x}=\left|V^\mu_{\bf x}\right|\tau^\mu$ called the \textit{PEE (thread) flow}. Here $\tau^\mu$ is the unit norm vector field that is tangent to the geodesics emanating from $\bf x$ everywhere in the bulk. The norm $\left|V^\mu_{x}\right|$ that characterizes the density of the PEE threads can be determined by the  requirement\footnote{See equation $\left(21\right)$ and Fig.$3$ in \cite{Lin:2023rxc} for more details.} that the number of PEE threads connecting two non-overlapping regions $A$ and $B$ equals to the PEE $\mathcal{I}\left(A,B\right)$. Note that, using vector fields to describe PEE threads introduces redundant information, which is the orientation of the threads. We want to stress that, our setup for the PEE threads are unoriented lines and for any given bulk surface we will calculate the number of intersections between the surface and the PEE threads, which is essentially different from the ``flux'' of the PEE threads passing through the surface. Hence, unlike \cite{Lin:2023rxc}, in this paper we avoid using the word ``PEE flux'' to prevent unnecessary confusion. 
	
	To be specific, in Poincar\'{e} AdS$_3$ with unit AdS radius,
		\begin{equation}
			ds^2=\frac{-dt^2+dz^2+dx^2}{z^2}
		\end{equation}
		the PEE thread flow $V_x^\mu$ on a time slice was carried out in \cite{Lin:2023rxc},
		\begin{equation}
			V^\mu_{\bf x}\left(\bar{x},\bar{z}\right)=\frac{1}{4G}\frac{2\bar{z}^2\left(\bar{x}-x\right)}{\left(\bar{x}-x\right)^2+\bar{z}^2}\left(z,\frac{\bar{z}-\left(\bar{x}-x\right)^2}{2\left(\bar{x}-x\right)}\right)
		\end{equation}
	In this context, the RT formula for a boundary region $A$ is reproduced by considering a homologous surface $\Sigma_{A}$ in the bulk and counting the number of intersections between the bulk PEE threads (or the PEE network) and $\Sigma_{A}$. Then the RT surface is exactly the homologous surface that has the minimal number of intersections with the bulk PEE threads \cite{Lin:2024dho}. For any bulk codimension-two surface $\Sigma$ we denote the number of intersections with the PEE threads as $L\left(\Sigma\right)$. A PEE thread may intersect with $\Sigma_{A}$ for multiple times, and each time of intersection should give positive contribution to $L\left(\Sigma\right)$. This is different from computing the flux of a divergenceless flow passing through $\Sigma$, where the in-going and out-going flow cancel with each other.
	
	There are two equivalent ways to calculate $L\left(\Sigma\right)$ \cite{Lin:2023rxc,Lin:2024dho}. The first way is to count the number of intersections for each PEE threads, then we sum over the numbers for all the threads. More explicitly, given a $\Sigma$, we can read the number $\omega_{\Sigma}\left(\bf x,\bf y\right)$ of intersections for any PEE thread connecting the boundary points $\bf x$ and $\bf y$ with density $\mathcal{I}(\bf x,\bf y)$. Then the contribution from this PEE thread to $L\left(\Sigma\right)$ is just $\omega_{\Sigma}\left(\bf x,\bf y\right)\mathcal{I}\left(\bf x,\bf y\right) d \textbf{x} d \bf y $, and 
	\begin{equation}\label{flux count PEE threads}
		L\left(\Sigma\right)=\frac{1}{2}\int_{\partial \mathcal{M}}d \textbf{x}\int_{\partial \mathcal{M}} d \textbf{y}\:\omega_{\Sigma}\left(\bf x,\bf y\right)\mathcal{I}\left(\bf x,\bf y\right)\,.
	\end{equation}
	Here $\mathcal{M}$ denotes the bulk AdS space on a time slice and $\partial\mathcal{M}$ is the AdS boundary, and $\omega_{\Sigma}\left(\bf x,\bf y\right)$ is also called the weight of the PEE thread connecting the boundary points $\bf x$ and $\bf y$. The factor $1/2$ appears due to double-counting, i.e. both $\mathcal{I}\left(\bf x, \bf y\right)$ and $\mathcal{I}\left(\bf y, \bf x\right)$ are counted in the integration. The second way is to compute $L\left(\Sigma\right)$ by counting the number of intersections along the surface using the vector field $V_x^\mu$. We integrate the density of the number of intersections along the surface $\Sigma$. This can be achieved by using the vector field $V_x^\mu $ without orientation. More explicitly we have,
	\begin{equation}\label{flux with flow vector}
		L\left(\Sigma\right)=\frac{1}{2}\int_{\Sigma}d\Sigma \sqrt{h}\int_{\partial \mathcal{M}}d\textbf{x}\left|V_{\textbf{x}}^\mu n_\mu\right|
	\end{equation} 
	where $h$ is the induced metric on $\Sigma$ and $n^\mu$ is its unit normal vector. Similarly, the coefficient $1/2$ appears as we doubly count each PEE thread by integrating $\bf x$ over the whole boundary $\partial\mathcal{M}$. And the reason why the absolute value is taken is that any PEE thread passing through $\Sigma$ gives positive contribution, no matter from which side it passes through $\Sigma$.
	
\begin{figure}
		\centering
		\begin{tikzpicture}[scale=1]
			\clip (-6,-1.5) rectangle (6,5);
			\filldraw[LouisBlue!30](-4,0) arc (180:0:4);
			\filldraw[LouisBlue!30](1,0) arc (180:0:1);
			\filldraw[white](-1,0) arc (180:0:1);
			\draw[ultra thick,LouisBlue] (-4,0) arc (180:0:4) ;
			\draw[ultra thick,LouisBlue] (-1,0) arc (180:0:1) ;
			\draw[very thick] (-9,0) -- (9,0);
			\draw[] (0,-0.4) node {$B_1$};
			\draw[] (2.5,-0.4) node {$A_2$};
			\draw[] (-2.5,-0.4) node {$A_1$};
			\draw[] (-5,-0.4) node {$B_2$};
			\draw[] (5,-0.4) node {$B_2$};
			\draw[ thick,dashed,LouisColor1] (-2,0) arc (180:0:2) ;
			\draw[ thick,dashed,LouisOrange] (0,0) arc (180:0:2.5) ;
			\draw[ thick,dashed,LouisColor2] (-3,0) arc (0:180:1) ;
			\draw[ thick,dashed,LouisOrange] (-5.7,3) -- (-5,3) ;
			\draw[] (-4.3,3.05) node {$\omega=2$};
			\draw[ thick,dashed,LouisColor1] (-5.7,2.) -- (-5,2.) ;
			\draw[] (-4.3,2.05) node {$\omega=0$};
			\draw[thick,dashed,LouisColor2] (-5.7,2.5) -- (-5,2.5) ;
			\draw[] (-4.3,2.55) node {$\omega=1$};
		\end{tikzpicture}
		\caption{The figure is extracted from \cite{Lin:2023rxc}. Here we consider a two-interval $A=A_1\cup A_2$ whose entanglement wedge is disconnected. The blue line is the RT surface $\mathcal{E}_{A}$ and the dashed lines are bulk PEE threads. One can see that the PEE threads (red) connectin $B_1$ and $B_2$ intersect with $\mathcal{E}_{A}$ twice hence giving a weight $\omega=2$.
		}
		\label{fig:twointerval}
	\end{figure}
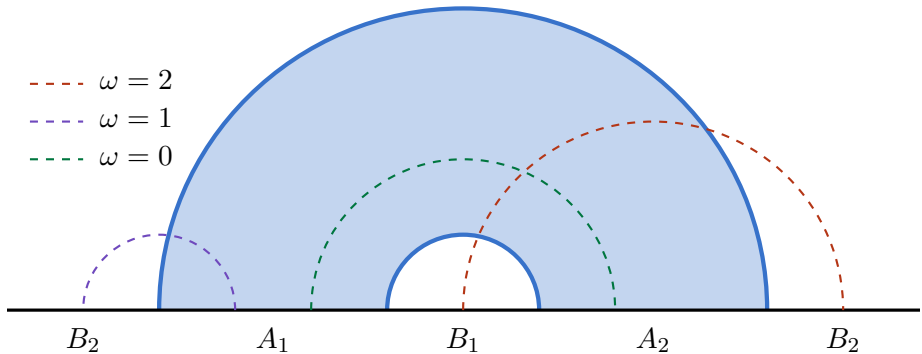

	The key observation in \cite{Lin:2024dho} is that, for the vacuum state dual to Poincar\'e AdS in general dimensions we have
	    \begin{equation}\label{any point any direction}
		\frac{1}{2}\int_{\partial \mathcal{M}}d\textbf{x}\left|V_{\bf x}^\mu n_\mu\right|=\frac{1}{4G}\,,
	\end{equation}
	which means that, the density of intersections at any point on any bulk surface $\Sigma$ is given by the constant $1/4G$. In other words, the AdS space is covered by the PEE threads uniformly everywhere. After substituting into \eqref{flux with flow vector}, we can find that the number of intersections $L\left(\Sigma\right)$ between any $\Sigma$ and the PEE network is nothing but
	\begin{equation}\label{fluxSigma}
		L\left(\Sigma\right)=\frac{\text{Area}\left(\Sigma\right)}{4G}\,.
	\end{equation}
	Note that the above equation hold for an arbitrary bulk surface, which may not be homologous to any boundary region.	Then the RT formula can be reproduced by minimizing the number of intersections $L\left(\Sigma_{A}\right)$ among all possible surfaces $\Sigma_{A}$ homologous to the boundary region $A$, i.e. \cite{Lin:2024dho}
	\begin{align}\label{equation: RT formula reformulation}
		S_A=&\min_{\Sigma_A}L\left(\Sigma_A\right)=\min_{\Sigma_A}\frac{1}{2}\int_{\Sigma_A}d\Sigma_A\sqrt{h}\int_{\partial \mathcal{M}}dx \left|V_x^\mu n_\mu\right|
		\cr
		=&\min_{\Sigma_A}\frac{\text{Area}\left(\Sigma_{A}\right)}{4G}
		\cr
		=&\frac{\text{Area}\left(\mathcal{E}_{A}\right)}{4G}
	\end{align}
	where $\mathcal{E}_{A}$ is exactly the RT surface. This gives us a very concrete geometric description on the contribution to a holographic entanglement entropy $S_{A}$ from each bulk PEE thread. More explicitly, only the two-point PEEs, whose PEE thread intersects with the RT surface $\mathcal{E}_{A}$, give non-zero contribution to $S_{A}$, and the amount of contribution equals the number it intersects with $\mathcal{E}_{A}$ \cite{Lin:2023rxc,Lin:2024dho}. See Fig.\ref{fig:twointerval} for an example of two disconnected intervals.
	
	A crucial implication of the above prescription to compute the holographic entanglement entropy is that, the naive normalization property of PEE: $\mathcal{I}(A,\bar{A})=S_{A}$ does not hold in general. In \cite{Lin:2023rxc} it was demonstrated that, the normalization property only applies to spherical boundary regions in Poincar\'e AdS, since in these configurations only the PEE threads connecting $A$ and $\bar{A}$ intersect $\mathcal{E}_{A}$ with weight $\omega=1$. This means we can define the contribution to $S_{A}$ from any subregion $\alpha$ of $A$, which is just the number of PEE threads emanating from $\alpha$ and ending on $\bar{A}$, i.e.
	\begin{align}
		s_{A}(\alpha)=\int_{\alpha}dx\int_{\bar{A}}dy\:\mathcal{I}\left(x,y\right)=\mathcal{I}(\alpha,\bar{A})\,,\qquad \text{for spherical regions,}
	\end{align} 
	which demonstrates the equivalence between the contribution representation and two-body correlation representation \eqref{tworep}. Nevertheless, this simple picture is no longer true for any non-spherical boundary regions. A typical example is the two-interval case shown in Fig.\ref{fig:twointerval}, where we can see that all the PEE threads connectng $B_1$ and $B_2$ intersect with $\mathcal{E}_{A}$ for two times, which means they contribute doubly to $S_{A}$. This indicates
	\begin{align}
		S_{A}=\mathcal{I}(A,\bar{A})+2\mathcal{I}(B_1,B_2)\neq \mathcal{I}(A,\bar{A})
	\end{align}
	hence the normalization property breaks down. Also the number of PEE threads emanating from $\alpha$ and ending on $\bar{A}$ can not fully capture the contribution from $\alpha$ to $S_{A}$, hence the contribution representation and the two-body correlation representation are no longer equivalent to each other,
	\begin{align}
		s_{A}(\alpha)\neq\mathcal{I}(\alpha,\bar{A}),\qquad \text{for non-spherical regions.}
	\end{align}
	
	In early literatures about PEE (see \cite{Wen:2018whg,Wen:2019iyq,Wen:2020ech,Wen:2021qgx,Han:2019scu} for examples), the confugrations under consideration mainly focused on single intervals or spherical regions, and the normalization property or the equivalence of the two representations was taken for granted. Even in the recent paper \cite{Basu:2023wmv} on the PEE and the BPE in island phases, \eqref{tworep} was used to derive the generlized ALC proposal and calculate the BPE. Nevertheless, later we will propose a new configuration for the PEE threads in island phases and show that \eqref{tworep} does not hold in general for the same reason. Based on this new setup, we will explicitly compute the contribution representation $s_{A}(\alpha)$ and two-body correlation representation $\mathcal{I}(\alpha,\bar{A})$ of the PEE in island phases. Also we will give a demonstration for the two basic proposals which relates the two-point and four-point functions of twist operators to PEE.
	
	In the rest of this paper, we refer to the two-body correlation representation $\mathcal{I}(A,B)$ whenever we mention PEE, and we will call $s_{A}(\alpha)$ the \textit{entanglement contribution} instead.

	\section{Setup for island phase}\label{sec: Weyl CFT}
	In this section we introduce a simulation of the KR braneworld or AdS/BCFT correspondence based on holographic CFT$_2$ with a special type of Weyl transformations that introduces finite cutoff to certain region, where entanglement islands can emerge \cite{Basu:2022crn,Basu:2023wmv,Lin:2023ajt,Chandra:2024bkn}.  Following \cite{Chandra:2024bkn}, we take the scalar field which characterizes the Weyl transformation as a dynamical field. The scalar field actually determines the geometry of the CFT after the Weyl transformation, hence it is described by a gravitational theory, which is equivalent to a Liouville theory \cite{Caputa:2017urj,Polyakov:1981rd}. It was found that the saddle point of the Liouville action exactly gives us an AdS$_2$ geometry. In other words, the Weyl transformed holographic CFT$_2$ gives us an explicit realization of the model that consist of an AdS$_2$ gravity coupled to a CFT$_2$ bath, which is a common context where entanglement islands emerge \cite{Chandra:2024bkn}. So we can apply the island formula \eqref{island formula 1} to this Weyl transformed CFT$_{2}$. On the gravity side, the finite cutoff configuration pushes the cutoff points on the RT surface deep into the bulk to form a cutoff brane, which plays the role of the End of World (EoW) brane in the KR braneworld or AdS/BCFT configuration. In this section we will establish the simulation step by step.
	
	\subsection{Holographic Weyl transformed CFT$_{2}$}\label{sec: cutoff sphere}	
	Let us consider the vacuum state of a holographic CFT$_2$ on a Euclidean flat space as our start, the metric is 
	\begin{equation}\label{equation:flat space metric}
		ds^2=\frac{1}{\epsilon^2}\left(d\tau^2+dx^2\right)\,,
	\end{equation}
	where $\epsilon$ is an infinitesimal constant representing the UV cutoff of the boundary CFT, and it can be referred to as the boundary metric of its dual Euclidean AdS$_3$ spacetime in Poincar\'{e} patch,
	\begin{equation}\label{equation: dual metric}
		ds^2=\frac{1}{z^2}\left(dz^2+d\tau^2+dx^2\right)\,,
	\end{equation}
	with the cutoff settled to be $z = \epsilon$. One may apply the Weyl transformation to the metric \eqref{equation:flat space metric},
	\begin{equation}\label{WeylTrans}
		ds^2=e^{2\phi\left(x\right)}\left(\frac{d\tau^2+dx^2}{\epsilon^2}\right)\,,
	\end{equation} 
	which effectively changes the cutoff scale as
	\begin{equation}\label{cutoff changes}
		\epsilon\Rightarrow e^{-\phi\left(x\right)} \epsilon\,.
	\end{equation}
	The Weyl transformations that are interesting to us are those amplify the cutoff scale, hence $\phi(x)$ is negative or zero for any $x$ .
	
	The two-point function of twist operators at the endpoints of an arbitrary single interval $A=\left[a,b\right]$ after the Weyl transformation picks up additional contributions from the scalar field $\phi\left(x\right)$ as follows \cite{Camargo:2022mme,Caputa:2017urj,Caputa:2018xuf},
	\begin{equation}\label{entropy under Weyl transformation}
		\tilde{S}_{[a,b]}=\frac{c}{3}\log\left(\frac{b-a}{\epsilon}\right)-\frac{c}{6}|\phi\left(a\right)|-\frac{c}{6}|\phi\left(b\right)|
	\end{equation}
	It is worth noting that, the above result \eqref{entropy under Weyl transformation} should not be understood as the entanglement entropy of $A$ as in the Weyl transformed CFT$_2$, due to the emergence of island configurations. This implies that the degrees of freedom in the Weyl transformed CFT$_{2}$ are not all independent, hence the formula for entanglement entropy should change (see \cite{Basu:2022crn} for more discussion). In \cite{Basu:2023wmv} the authors gave two basic proposals to interpret these two-point functions in terms of PEE (see also section \ref{sec3.3}). 
	
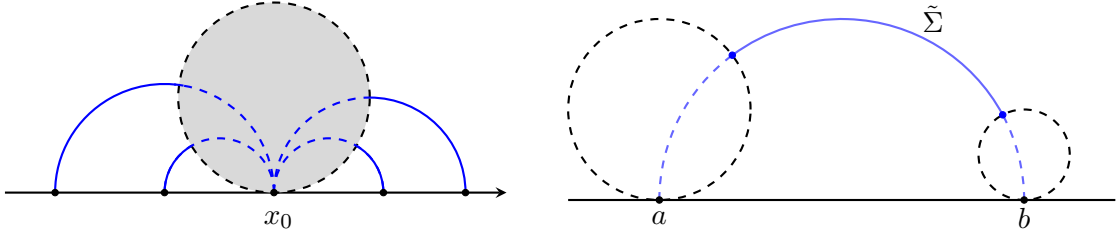
\begin{figure}
		\centering
		\begin{tikzpicture}[scale=1.2]
			\draw[thick,->,>=stealth]
			(-5.5,0)--(0,0);
			\draw[thick,dashed,fill=black!15,opacity=40]
			(-1.5,1.05) arc (0:360:1.05);
			\draw[thick,dashed,blue]
			(-2.55,0) arc (180:90:1.05);
			\draw[thick,blue]
			(-0.45,0) arc (0:90:1.05);
			\draw[thick,dashed,blue]
			(-2.55,0) arc (0:85:1.2);
			\draw[thick,blue]
			(-4.95,0) arc (180:80:1.2);
			\draw[thick,dashed,blue]
			(-2.55,0) arc (180:0:0.6);
			\draw[thick,blue]
			(-1.35,0) arc (0:60:0.6);
			\draw[thick,dashed,blue]
			(-2.55,0) arc (0:180:0.6);
			\draw[thick,blue]
			(-3.75,0) arc (180:120:0.6);
			\filldraw[black] (-2.55,0) circle (1pt);
			\filldraw[black] (-0.45,0) circle (1pt);
			\filldraw[black] (-4.95,0) circle (1pt);
			\filldraw[black] (-3.75,0) circle (1pt);
			\filldraw[black] (-1.35,0) circle (1pt);
			\draw[] (-2.5,-0.3) node {$x_0$};
		\end{tikzpicture}
		\hspace*{15pt}
		\begin{tikzpicture}[scale=1.2]
			\draw[thick]
			(-3,0)--(3,0);
			\draw[thick,black,dashed]
			(2.5,0.5) arc (0:360:0.5);
			\draw[thick,black,dashed]
			(-1,1) arc (0:360:1);
			\draw[thick,blue!60]
			(1.765,0.941) arc (28.08:126.87:2);
			\draw[thick,blue!60,dashed]
			(2,0) arc (0:28.08:2);
			\draw[thick,blue!60,dashed]
			(-2,0) arc (180:126.87:2);
			\filldraw[black] (2,0) circle (1pt);
			\filldraw[black] (-2,0) circle (1pt);
			\filldraw[blue] (1.765,0.941) circle (1pt);
			\filldraw[blue] (-1.2,1.6) circle (1pt);
			\draw[] (2,-0.2) node {$b$};
			\draw[] (-2,-0.2) node {$a$};
			\draw[] (1,2) node {$\tilde{\Sigma}$};
		\end{tikzpicture}
		\caption{Left: The gray region is a cutoff sphere at $x=x_0$ in the bulk dual geometry for a Weyl transformed CFT. For geodesics anchored at $x_0$ represented by blue lines, the portion inside the cutoff sphere has length $\abs{\phi(x_0)}$ which is excluded; Right: In the Weyl transformed CFT, the RT surface $\tilde{\Sigma}$ (blue line) of interval $\left[a,b\right]$ is cut by two cutoff spheres (black dashed line) at the two endpoints, with the inside lengths (blue dashed lines) are exactly $\abs{\phi(a)},\abs{\phi(b)}$ at $x=a,x=b$ respectively.}
		\label{fig:cutoff-sphere}
	\end{figure}

	Holographically, this formula can be understood more easily through the perspective of the alternation of cutoff by introducing the \emph{cutoff sphere} \cite{Basu:2022crn} to manifest the adjustment of cutoff points on the RT surface. To be specific, consider a time slice of AdS$_3$ and any boundary point $\left(x_0,\epsilon\right)$, a general Weyl transformation results in inserting a cutoff sphere tangent at this boundary point with the radius determined by $|\phi(x_0)|$, which is described by \cite{Chandra:2024bkn},
	\begin{equation}
		\left(x-x_0\right)^2+\left(z-\alpha\right)^2=\alpha^2,\quad \alpha\equiv \frac{\epsilon}{2}e^{\abs{\phi(x_0)}}\,.
	\end{equation}
	It is interesting to note that, the above equation also describes a circle in the flat background with the center being $\left(x_0,\alpha\right)$ and the radius $\alpha$. This is also noted as the \textit{horocycle} in the literature (see \cite{Levay:2019nsr} for example). The cutoff sphere is defined such that, when we integrate over \emph{any} RT surfaces anchored at $\left(x_0,\epsilon\right)$, we only integrate up to the cutoff sphere and the part inside the cutoff sphere is abandoned. The length of the abandoned part is exactly the constant $\abs{\phi(x_0)}$, see the left figure of Fig.\ref{fig:cutoff-sphere}. What follows is that, the RT surface of $A=\left[a,b\right]$ in the Weyl transformed CFT are cut by two cutoff spheres at the endpoints, which reduces the length of the RT surface by $\abs{\phi(a)},\abs{\phi(b)}$ respectively, as shown in the right figure of Fig.\ref{fig:cutoff-sphere}. As a consequence, such reduction of length explains the difference between \eqref{entropy under Weyl transformation} and $S_A$ in CFT$_2$. Nevertheless this holographic description for \eqref{entropy under Weyl transformation} is just an observation with no physical demonstration. Later we will come back to this point.

	\subsection{A simulation of KR braneworld or AdS/BCFT by Weyl transformation}
    The key requirement that the Weyl transformed CFT$_2$ is a simulation of the KR braneworld or AdS/BCFT configurations is that, the cutoff brane induced by the Weyl transformation should match to the KR brane, which was adjusted by hand in \cite{Basu:2022crn,Lin:2023ajt}. Later it was found in \cite{Chandra:2024bkn} that, there is an intrinsic reason to determine the Weyl transformation such that the corresponding cutoff brane match to the EoW branes. More explicitly, the Weyl transformations under consideration are those that optimize the path integral computation for the reduced density matrix of an interval in holographic CFT$_2$.
    By \emph{path integral optimization} we mean that, this special Weyl transformation minimizes the path integral complexity and preserves the reduced density matrix $\rho_A$ for a given interval $A$ at a particular time in CFT$_2$ \cite{Caputa:2017urj,Caputa:2017yrh}. 
    
    To be specific, consider the wave function of the ground state at $\tau=-\epsilon$ in the Euclidean CFT$_2$:
	\begin{equation}\label{wave function}
		\Psi_{\delta_{a b} / \epsilon^2}[\tilde{\varphi}(\xi)]=\int\left(\prod_{\xi} \prod_{-\infty<\tau<-\epsilon} D \varphi(\tau, \xi)\right) e^{-S_{C F T}(\varphi)} \cdot \prod_{\xi} \delta(\varphi(-\epsilon, \xi)-\tilde{\varphi}(\xi)) .
	\end{equation}
	where $\delta_{a b}=d\tau^2+d\xi^2$ and the subscript of $\Psi$ is to denote the metric where the path integral is performed. Let us conduct a general Weyl transformation \eqref{WeylTrans} additionally with a boundary condition for the scalar field $\phi$:
	\begin{equation}
		d s^2=e^{2 \phi(\tau, \xi)} \frac{d \tau^2+d \xi^2}{\epsilon^2}, \quad e^{2 \phi(\tau=-\epsilon, \xi)}=1
	\end{equation}
	Then the wave function $\Psi$ evaluated under the above Weyl transformation is proportional to \eqref{wave function},
	\begin{equation}
		\Psi_{e^{2 \phi} \delta_{a b} / \epsilon^2}[(\tilde{\varphi}(\xi))]=e^{C_L[\phi]-C_L[0]} \Psi_{\delta_{a b} / \epsilon^2}[(\tilde{\varphi}(\xi))],
	\end{equation}
	which means the state $\Psi$ is preserved under the Weyl transformation, and the Liouville action $C_L[\phi]$ is given by \cite{Polyakov:1981rd},
	\begin{equation}
		C_L[\phi]=\frac{c}{24 \pi} \int_{-\infty}^{\infty} d \xi \int_{-\infty}^{-\epsilon} d \tau\left(\left(\partial_{\xi} \phi\right)^2+\left(\partial_\tau \phi\right)^2+\mu e^{2 \phi}\right)
	\end{equation}
	which is also referred to as the \emph{path integral complexity} of the state $\Psi$ \cite{Caputa:2017urj,Caputa:2017yrh}. Then the path integral optimization which minimize the path integral complexity is achieved by solving the equation of motion $(\partial_\xi^2+\partial_\tau^2)\phi=e^{2\phi}/\epsilon^2$:
	\begin{equation}\label{eq: path integral complexity eq}
		e^{2 \phi}=\frac{\epsilon^2}{\tau^2}
	\end{equation}
	
	The above procedure optimizes the path integral for the whole time slice in Euclidean CFT$_2$. For the path integral optimization works for a single interval $A=[a,b]$, one can work in the complex coordinate $(\eta, \bar{\eta})=(x+i t, x-i t)$ and use the following conformal transformation
	\begin{equation}
		w=\sqrt{\frac{\eta-a}{b-\eta}}, \quad(\omega, \bar{\omega})=(\xi+i \tau, \xi-i \tau) .
	\end{equation}
	to maps the interval $A$ to the whole infinite line to which \eqref{eq: path integral complexity eq} can apply. Therefore, by requiring that
	\begin{equation}
		\frac{\epsilon^2}{\tau^2} d \omega d \bar{\omega}=e^{2 \phi(x)} d \eta d \bar{\eta}
	\end{equation}
	which implies,
	\begin{equation}
		\phi\left(x\right)=\left\{\begin{array}{cl}
			0, & a<x<b \\
			\log \left[\frac{\epsilon(b-a)}{2(x-a)(x-b)}\right]+\kappa, & x>b \text { or } x<a
		\end{array}\right.
	\end{equation}
    
    Now we take the limit $a\rightarrow 0$ and $b\rightarrow\infty$. In other words, we optimizes the path integral computation for the reduced density matrix $\rho_A$ of the right half-plane $A=[0,\infty)$. Then, we have a special UV-cutoff-dependent Weyl transformation for the $x<0$ region as our setup to eliminate the short-range entanglement: 
\begin{equation}\label{scalar field}
	\phi\left(x\right)=\left\{
	\begin{aligned}
		&0\quad &\text{if }\quad  x\geq 0\\
		&-\log\left(\frac{2\left|  x \right|}{\epsilon}\right)+\kappa ,\quad &\text{if }\quad  x<0
	\end{aligned}
	\right.
\end{equation} 
where $\kappa$ is an undetermined constant. After the transformation, the metric \eqref{WeylTrans} in $x<0$ becomes
\begin{equation}\label{setup}
	ds^2=\frac{e^{2\kappa}}{4}\left(\frac{d\tau^2+dx^2}{x^2}\right),\quad x<0
\end{equation}
such that the metric in $x<0$ region is proportional to the AdS$_2$ metric. The specific Weyl transformation \eqref{scalar field} results in a cutoff sphere for the boundary point $\left(x_0,\epsilon\right)$ where $x_0<0$ \cite{Basu:2022crn},
	\begin{equation}
		\left(x-x_0\right)^2+\left(z-\left|x_0\right|e^{-\kappa}\right)^2=\left|x_0\right|^2e^{-2\kappa}
	\end{equation}
	with the center being $\left(x_0,\left|x_0\right|e^{-\kappa}\right)$ and the radius
	\begin{equation}\label{radius}
		r\equiv\left|x_0\right|e^{-\kappa}.
	\end{equation}
Meanwhile, there exists a common tangent line of all cutoff spheres in $x<0$ region, which we call the \emph{cutoff brane} \cite{Basu:2022crn}, and it plays the same role as the Karch-Randall (KR) brane (or End of World (EoW) brane) in the KR braneworld or AdS/BCFT configuration (see Fig.\ref{fig:the-common-tangent-line}), indicating that the entanglement structures of this Weyl transformed CFT and the KR braneworld are similar, see \cite{Basu:2022crn} for more details.
	
Note that, when we optimize the path-integral under Weyl transformations, we are treating the scalar field $\phi\left(x\right)$ as a dynamical field with the action being the Liouville action. More interestingly, since $\phi\left(x\right)$ is a metric factor, the Liouville action should be a gravitational theory\footnote{See \cite{Coussaert:1995zp,Levay:2019nsr} for earlier discussions about the relations between Louville theory and three-dimensional gravity.} with AdS$_2$ geometric saddle points \eqref{setup}. In summary, we arrive at a setup where an AdS$_2$ gravity in the $x<0$ region is coupled to a non-gravitational CFT$_2$ in the $x>0$ region, to which the island formula \eqref{island formula 1} applies.

For example, let us consider the interval $A$ to be the semi-infinite region $x>a$ and take the region $I=\left(-\infty,-a'\right]$ as a possible island choice for $A$. According to the island formula, the entanglement entropy $S_{A}$ is given by\footnote{Note that we have ignored the area term since the gravitational theory is an induced gravity.},
		\begin{equation}\label{shexian}
			S_A=\min_{I}\{\tilde{S}_{A\cup I}\}=\min_{a'}\left\{\frac{c}{3}\log\left(\frac{a+a'}{\epsilon}\right)-\frac{c}{6}\log\left(\frac{2a'}{\epsilon}\right)+\frac{c}{6}\kappa\right\}
		\end{equation} 
		One can easily check $a'=a$ is the only minimal value of above equation,
		\begin{equation}
			S_A=\frac{c}{6}\log\left(\frac{2a}{\epsilon}\right)+\frac{c}{6}\kappa
		\end{equation}
		The result is visualized in Fig.\ref{fig:rt-cutoff-surface} in the bulk, where the RT surface $\tilde{\Sigma}$ is a semi-circle with one endpoint normally anchored on the cutoff sphere of $x=-a$, i.e.  normally anchored on the cutoff brane,
		\begin{equation}\label{RT surface}
			\tilde{\Sigma}=\left\{\left(x,z\right)|z=\sqrt{a^2-x^2},z\geq qx\right\}
		\end{equation}
		where $q$ is the slope of the cutoff brane. Similar discussions hold for more complicated configuration.

			\begin{figure}
			\centering
			\begin{tikzpicture}[scale=1.0]
				\draw[thick]
				(-5,0)--(5,0);
				\draw
				(-1.980,2.263)--(-1.838,2.121)--(-1.980,1.980);
				\path[fill=blue!20,opacity=80]
				(5,0) --(5,5)--(-5,5)--(-2.121,2.121) arc (135:0:3)--cycle ;
				\draw[thick,orange]
				(3,0) arc (0:135:3);
				\draw[thick,dashed,orange]
				(3,0) arc (0:180:3);
				\draw[thick,black,dashed] (-3,1.243) circle (1.243);
				\draw[thick,red]
				(0,0)--(-5,5);
				\filldraw[red] (3,0) circle (1pt);
				\filldraw[red] (-3,0) circle (1pt);
				\draw[thick,red]
				(3,0)--(5,0);
				\draw[thick,red]
				(-5,0)--(-3,0);
				\draw[] (3,-0.3) node {$a$};
				\draw[] (-3,-0.3) node {$-a$};
				\draw[] (4,-0.3) node {$A$};
				\draw[] (-4.2,-0.3) node {Island};
				\draw[] (1.5,2.3) node {\textcolor{orange}{$\tilde{\Sigma}$}};
				\draw[] (3.5,3.5) node {\textcolor{blue!50}{$\mathcal{W}_A$}};
				\draw[] (-5,5.3) node {\textcolor{red}{cutoff brane}};
				\draw[thick,->,>=stealth]
				(6,1)--(6.5,1);
				\draw[thick,->,>=stealth]
				(6,1)--(6,1.5);
				\draw[] (6.5,0.8) node {$x$};
				\draw[] (5.8,1.5) node {$z$};
			\end{tikzpicture}
			\caption{The RT surface of region $A$ is represented by the orange line and the cutoff brane starting from $x=0$ is represented by the red line. The dashed black sphere is the cutoff sphere at $x=-a$. The entanglement wedge of $A$ denoted by $W_A$ is shaded in purple.}
			\label{fig:rt-cutoff-surface}
		\end{figure}
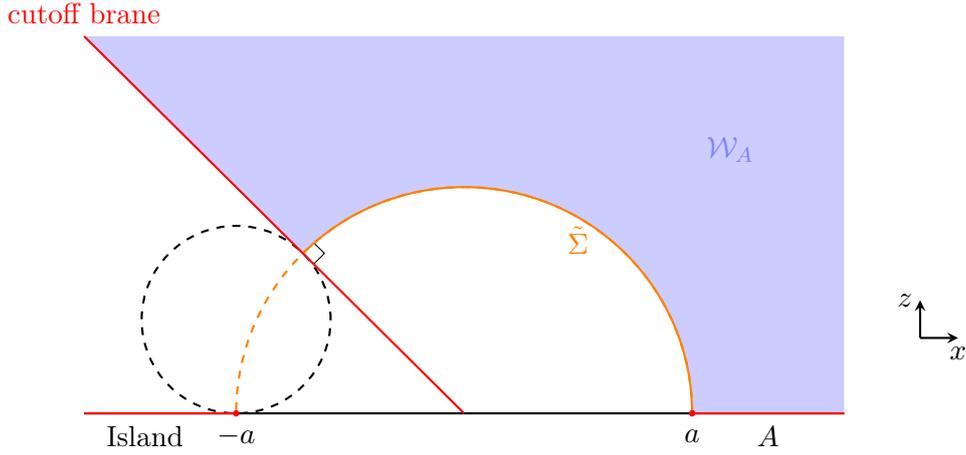
			\begin{figure}
			\centering
			\begin{tikzpicture}[scale=0.8]
				\draw[thick,->,>=stealth]
				(-8,0)--(8,0);
				\draw[thick,->,>=stealth]
				(0,0)--(0,7);
				\draw[thick,red]
				(0,0)--(-6,6);
				\draw[] (-6,6.2) node {\textcolor{red}{cutoff brane}};
				\draw[thick,blue!60]
				(2,0) arc (0:135:2);
				\draw[thick,blue!60,dashed]
				(-2,0) arc (180:135:2);
				\draw[thick,black,dashed]
				(-1.172,0.828) arc (0:360:0.828);
				\draw[thick,blue!60]
				(3,0) arc (0:135:3);
				\draw[thick,blue!60,dashed]
				(-3,0) arc (180:135:3);
				\draw[thick,black,dashed]
				(-1.757,1.243) arc (0:360:1.243);
				\draw[thick,blue!60]
				(4.5,0) arc (0:135:4.5);
				\draw[thick,blue!60,dashed]
				(-4.5,0) arc (180:135:4.5);
				\draw[thick,black,dashed]
				(-2.636,1.864) arc (0:360:1.864);
				\draw[thick,blue!60]
				(6.5,0) arc (0:135:6.5);
				\draw[thick,blue!60,dashed]
				(-6.5,0) arc (180:135:6.5);
				\draw[thick,black,dashed]
				(-3.808,2.692) arc (0:360:2.692);
				\filldraw[black] (0,0) circle (1pt);
				\filldraw[black] (2,0) circle (1pt);
				\filldraw[black] (3,0) circle (1pt);
				\filldraw[black] (4.5,0) circle (1pt);
				\filldraw[black] (6.5,0) circle (1pt);
				\filldraw[black] (-2,0) circle (1pt);
				\filldraw[black] (-3,0) circle (1pt);
				\filldraw[black] (-4.5,0) circle (1pt);
				\filldraw[black] (-6.5,0) circle (1pt);
				\filldraw[blue] (-1.414,1.414) circle (1pt);
				\filldraw[blue] (-2.121,2.121) circle (1pt);
				\filldraw[blue] (-3.182,3.182) circle (1pt);
				\filldraw[blue] (-4.596,4.596) circle (1pt);
				\draw[] (0,-0.2) node {$0$};
				\draw[] (8,-0.2) node {$x$};
				\draw[] (-0.2,7) node {$z$};
			\end{tikzpicture}
			\caption{The common tangent surface to all the cutoff spheres, which is called the cutoff brane represented by the red line, can be compared to the KR brane (or EoW brane) in the KR braneworld or AdS/BCFT configuration.}
			\label{fig:the-common-tangent-line}
		\end{figure}
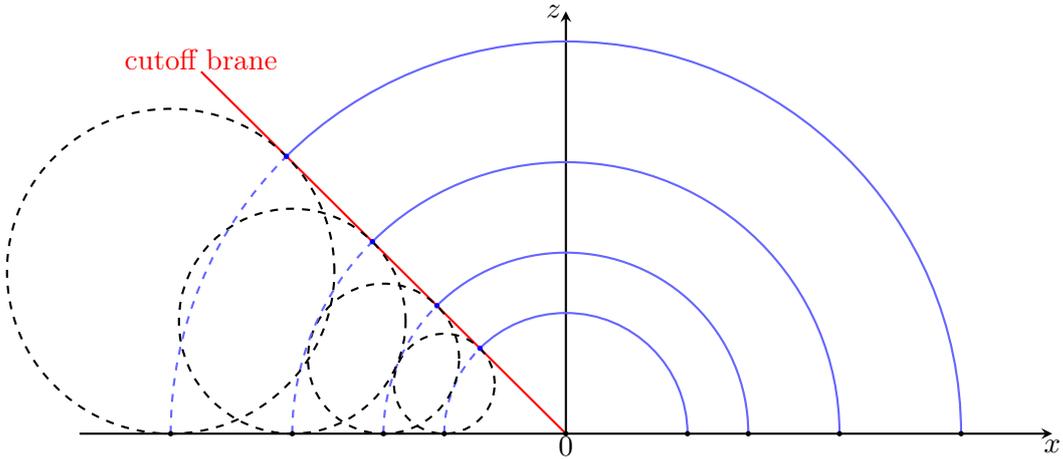
		
		\subsection{Proposals for PEE and correlation functions of twist operators in island phase}\label{sec3.3}
		The PEE structure in island phases of the Weyl transformed CFT$_2$ is actually not clear as in the holographic CFT$_2$. In \cite{Basu:2023wmv} the authors give two proposals that relate the PEE in Weyl transformed CFT$_2$ to the two-point functions of the twist operators \eqref{two-point PEE}. And based on these proposals, the authors generalized the concept of the balanced partial entanglement (BPE) to the island phases. Remarkably in \cite{Basu:2023wmv}, the correspondence between the BPE and the entanglement wedge cross section (EWCS) was explicitly checked in various phases of entanglement wedges in the AdS$_3$ bulk with a cutoff brane. We list the proposals in the following \cite{Basu:2023wmv}:
				\begin{itemize}
			\item Basic proposal $1$: When $\gamma$ is a single interval, we have
			\begin{equation}\label{Basic proposal 1}
				\tilde{S}_\gamma=\mathcal{I}\left(\gamma,\bar{\gamma}\right);
			\end{equation}
			\item Basic proposal $2$: For $A=\left[a,b\right]$ and its generalized island Ir$\left(A\right)=\left[-d,-c\right]$,
			\begin{equation}\label{Basic proposal 2}
				\tilde{S}_{A\text{Ir}\left(A\right)}=\tilde{S}_{\left[-d,b\right]}+\tilde{S}_{\left[-c,a\right]}.
			\end{equation}
		\end{itemize}
		Note that, in \cite{Basu:2023wmv} it was claimed that the basic proposal 2 can be derived from the basic proposal 1 such that it is not independent. Nevertheless, here we point out that the derivation in \cite{Basu:2023wmv} indeed used the basic proposal 2 (see equation (1.1) of \cite{Basu:2023wmv}), hence there exists a loop hole and the above two proposals are independent from each other. Provided that both of the above proposals are valid, we stress that the validity of the calculations and main results in \cite{Basu:2023wmv} are not affected by this loop hole. In section \ref{sec.PEEinislandphase} we will give support for the above proposals based on our new setup of PEE thread configurations in island phases.
		
		In proposal 1, $\tilde{S}_\gamma$ is just the two-point function of twist operators inserted at the endpoints of $\gamma$, which is consistent with \eqref{entropy under Weyl transformation}. The proposal 1 indicates that in island phase, any two-point function represents the PEE between the interval $\gamma$ enclosed by the two points and the complement $\bar{\gamma}$. In proposal 2, we consider two disconnected intervals $A$ and $\text{Ir}(A)$ and the four-point functions of twist operators settled at their endpoints. Here $A$ admits an island $\text{Is}(A)$ and $\text{Ir}(A)\supseteq \text{Is}(A)$ is called the generalized island (see section \ref{sec.revisited} for a brief introduction)
 of $A$, and the four-point function is denoted as $\tilde{S}_{A\text{Ir}(A)}$. Proposal 2 implies that the four-point function $\tilde{S}_{A\text{Ir}\left(A\right)}$ is given by the summation of two two-point functions, like the RT surface for two disconnected intervals with a connected entanglement wedge in AdS$_3$/CFT$_2$. This is a result of the large $c$ limit \cite{Hartman:2013mia,Faulkner:2013yia}.
		
Note that the correlation functions $\tilde{S}_\gamma$ and $\tilde{S}_{A\text{Ir}(A)}$ can only reproduce entanglement entropy when the optimization conditions in island formula \eqref{island formula 1} are satisfied, i.e.
\begin{align}
		&\tilde{S}_{\gamma}=S_{A},~~~~~ \quad when~~~~\gamma=A\cup \text{Is}(A)\,, \\
			&\tilde{S}_{A\text{Ir}(A)}= S_{A}, \quad when~~~~\text{Ir}(A)=\text{Is}(A)\,.
\end{align}
If we go beyond the above configurations, although the correlation functions are still well defined in the field theory side, their physical meaning in terms of entropy quantities are not well-understood in the literature. The two basic proposals give new interpretation for generic correlation functions in terms of PEEs, and give highly consistent results for the BPE which exactly reproduce the area of EWCS in various phases. Nevertheless the two basic proposals were not proved. In this paper we will give an interpretation for these proposals in the context of PEE network.

\section{PEE threads and correlation functions of twist operators in island phase}\label{sec.PEEinislandphase}
	
	\subsection{Setup for the PEE thread configuration}
	In this section we explore possible configurations of PEE threads, or PEE network, for the Weyl transformed CFT$_2$ where entanglement island emerges. The fundamental requirement is to reproduce the island formula \eqref{island formula 1} by optimizing the number of intersections between the PEE threads and certain homologous surface in the bulk. The introduction of cutoff spheres gives us a direction on how to construct the PEE network.
	
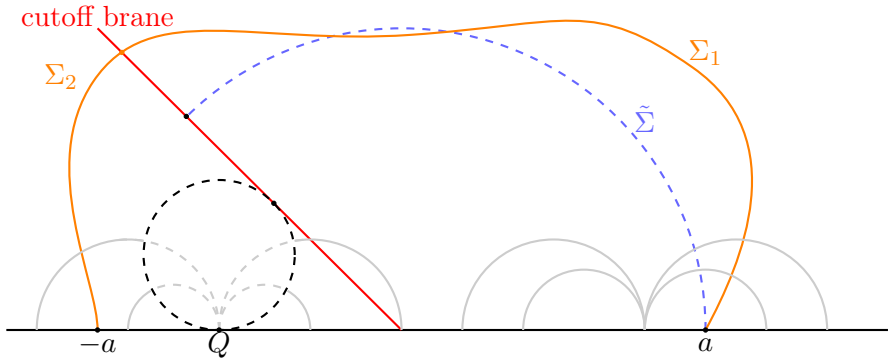
\begin{figure}
		\centering
		\begin{tikzpicture}[scale=0.8]
			\draw[thick]
			(-6.5,0)--(8,0);
			\draw[thick,red]
			(0,0)--(-5,5);
			\draw[] (-5,5.2) node {\textcolor{red}{cutoff brane}};
			\draw[thick,dashed,blue!60]
			(5,0) arc (0:135:5); 
			\draw[yscale=-1,xscale=1,thick,orange]    (-5,0) .. controls (-4.988,-0.7747) and (-6.0021,-2.7246) .. (-5.07,-4.1178) .. controls (-4.1379,-5.511) and (-2.4367,-4.7116) .. (0.1433,-4.8911) .. controls (2.7234,-5.0706) and (3.1433,-5.5178) .. (4.7967,-4.3444) .. controls (6.45,-3.1711) and (5.6133,-1.1338) .. (5,0) ;
			\draw[thick,gray!40]
			(4,0) arc (0:180:1); 
			\draw[thick,gray!40]
			(4,0) arc (0:180:1.5); 
			\draw[thick,gray!40]
			(4,0) arc (180:0:1); 
			\draw[thick,gray!40]
			(4,0) arc (180:0:1.5);
			\draw[thick,dashed]
			(-1.757,1.243) arc (0:360:1.243);  
			\draw[thick,gray!40]
			(-1.5,0) arc (0:60:0.75);  
			\draw[thick,gray!40,dashed]
			(-3,0) arc (180:60:0.75); 
			\draw[thick,gray!40]
			(0,0) arc (0:100:1.5);  
			\draw[thick,gray!40,dashed]
			(-3,0) arc (180:90:1.5);
			\draw[thick,gray!40]
			(-4.5,0) arc (180:120:0.75);  
			\draw[thick,gray!40,dashed]
			(-3,0) arc (0:120:0.75);
			\draw[thick,gray!40]
			(-6,0) arc (180:80:1.5);  
			\draw[thick,gray!40,dashed]
			(-3,0) arc (0:100:1.5);
			\filldraw[] (-3.54,3.54) circle (1pt);
			\filldraw[orange] (-4.6,4.6) circle (1pt);
			\filldraw[] (-2.1,2.1) circle (1pt);
			\filldraw[] (5,0) circle (1pt);
			\filldraw[] (-5,0) circle (1pt);
			\filldraw[] (-3,0) circle (1pt);
			\draw[] (-3,-0.25) node {$Q$};
			\draw[] (-5,-0.25) node {$-a$};
			\draw[] (5,-0.25) node {$a$};
			\draw[] (5,4.6) node {\textcolor{orange}{$\Sigma_1$}};
			\draw[] (-5.6,4.2) node {\textcolor{orange}{$\Sigma_2$}};
			\draw[] (4,3.5) node {\textcolor{blue!60}{$\tilde{\Sigma}$}};
		\end{tikzpicture}
		\caption{The orange solid line $\Sigma=\Sigma_{1}\cup\Sigma_{2}$ is one example of the homologous curves connecting $x=\pm a$, which is decomposed into two parts where $\Sigma_{1}$ is in the region beyond the cutoff brane while $\Sigma_{2}$ is behind the cutoff brane. The blue dashed line $\tilde{\Sigma}$ is the minimal surface connecting the boundary point at $x=a$ and the cutoff brane. A naive prescription is to assume that the PEE threads (gray lines) are cut off by the cutoff spheres (black dashed circles) such that PEE threads emanating from $Q$ are largely reduced but not empty, which fails to give the island formula.}
		\label{fig:errortrial}
	\end{figure}

	For a pair of boundary points at $x=\pm a$, according to the island formula \eqref{island formula 1} the two-point function $\tilde{S}_{[-a,a]}$ is given by the area of the homologous geodesic connecting these two points but cut off at the cutoff brane (see the curve $\tilde{\Sigma}$ in Fig.\ref{fig:rt-cutoff-surface}),
	\begin{align}
		S_{[0,a]}=\tilde{S}_{[-a,a]}=\frac{\text{Area}(\tilde{\Sigma})}{4G}\,.
	\end{align}
	One may naively consider the following way to modify the PEE network: for any boundary point $Q$ in the Weyl transformed region ($x<0$), the PEE threads emanating from $Q$ are terminated at the corresponding cutoff sphere associated to $Q$. This seems to be a natural step since the RT surfaces are cutoff at the cutoff spheres. In this setup, the PEE network beyond the cutoff brane is not changed, while in the region behind the cutoff brane the density of PEE threads is largely reduced but not empty (see Fig.\ref{fig:errortrial}). Now we consider all possible homologous curves $\Sigma$ connecting the $x=\pm a$ boundary points and minimize the number of intersections with the modified PEE network. We can always decompose $\Sigma$ into two parts $\Sigma=\Sigma_{1}\cup\Sigma_{2}$ where $\Sigma_{1}$ is in the region beyond the cutoff brane while $\Sigma_{2}$ is behind the cutoff brane. In the region beyond the brane, since the PEE network is not changed, according to \eqref{fluxSigma} the number of intersections for $\Sigma_{1}$ is given by
	\begin{equation}
		L(\Sigma_{1})=\frac{\text{Area}(\Sigma_{1})}{4G}\,.
	\end{equation}
Since $\tilde{\Sigma}$ is the minimal surface connecting the boundary point at $x=a$ and the cutoff brane, we have 
	\begin{equation}
	\text{Area}(\Sigma_{1})/4G\geq \text{Area}(\tilde{\Sigma})/4G\,.
	\end{equation} 
	In the region behind the brane, the number of intersections is positive as there still exits PEE threads behind the brane,
	\begin{equation}
		L(\Sigma_{2})>0\,.
	\end{equation}
	In summary, for any $\Sigma$ we always have
	\begin{align}
		L(\Sigma)=L(\Sigma_{1})+L(\Sigma_{2})>L(\tilde{\Sigma})=S_{[0,a]}\,,
	\end{align}
	which means that in this prescription we cannot reproduce the result of the island formula via optimizing the number of intersections.
	Although the PEE threads stretch along the bulk geodesics, which coincides with the RT surfaces, the PEE threads and the RT surface are distinct physical quantities. Hence, the setup that the RT surfaces are cut off at the cutoff spheres does not mean the same to the PEE threads, and we should not be surprised to see the failure of the above naive prescription. 
	
	In the holographic picture of the Weyl transformed CFT, it is always the RT surfaces that are cut off at the cutoff spheres. When evaluating the entropy quantities in terms of the PEE threads, instead of modifying the network of the PEE threads, we may consider modifying the configurations of the boundary points. In the following we give a new prescription to establish a configuration of the bulk PEE threads based on which we can evaluate entropy quantities via optimizing the number of intersections between PEE threads and all possible homologous surfaces. The prescription consists of the following key points:	
	\begin{enumerate}
		\item The PEE network for holographic Weyl transformed CFT$_2$ is the same as the network in holographic CFT$_2$.
		
		\item In holographic Weyl transformed CFT, the boundary points are replaced by the cutoff spheres determined by the value of the scalar field at this point. More explicitly, a boundary point $x=x_0$ is replaced by the cutoff sphere $\bigodot_{x_0}$ with the center $\left(x_0,\alpha\right)$ and radius $\alpha=\frac{\epsilon}{2}e^{\abs{\phi(x_0)}}$ \footnote{For boundary region where $\phi(x)=0$, the corresponding cutoff sphere has the center $(x,\epsilon/2)$ and radius $\epsilon/2$. The cutoff surface then becomes $z=\epsilon$, which is exactly where the RT surfaces are cut off for holographic CFT$_2$ without Weyl transformation.}.
		
		\item The surfaces that are homologous to any boundary interval $A=[a,b]$ are replaced by the homologous surfaces anchored on the corresponding cutoff spheres $\bigodot_{a}$ and $\bigodot_{b}$. Among all such possible homologous surfaces we will identify the one that has the minimized number of intersections with the PEE threads.
		
		\item In island configurations, when we calculate the entanglement entropy for a certain region, we should allow the homologous surface to anchor on the cutoff surfaces associated to \textit{any} boundary point such that optimizing the number of intersections amounts to finding the boundary point which gives the minimal homologous surface.
	\end{enumerate}
	
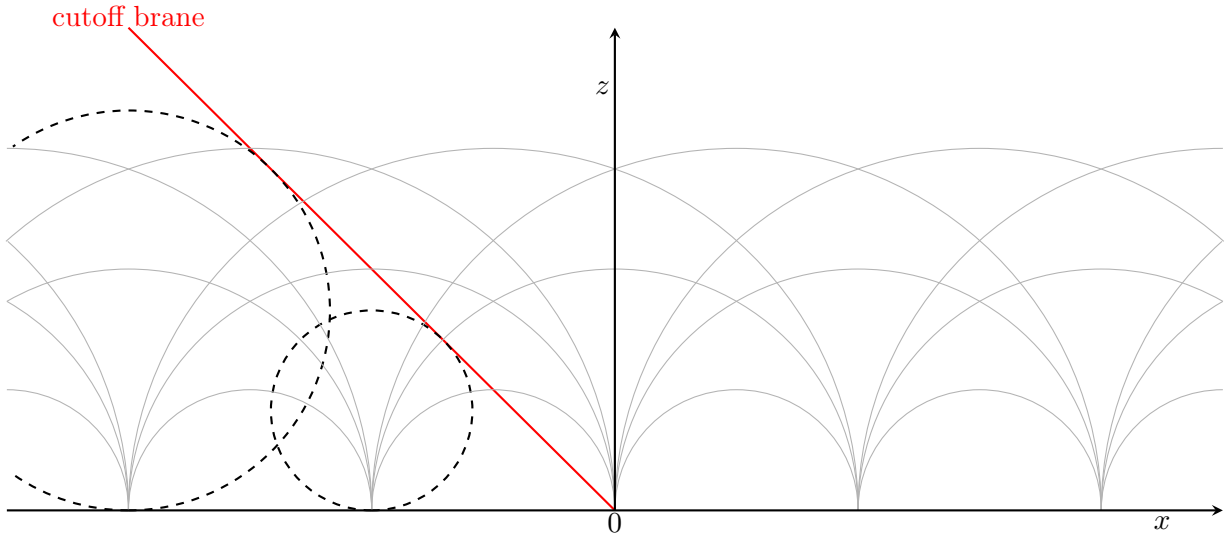
\begin{figure}
		\centering
		\begin{tikzpicture}[scale=0.8]
			\draw[thick,red]
			(0,0)--(-8,8);
			\draw[] (-8,8.2) node {\textcolor{red}{cutoff brane}};
			\draw[thick,black,dashed]
			(-4,0) arc (-90:270:1.657);
			\draw[thick,black,dashed]
			(-8,0) arc (-90:125:3.314);
			\draw[thick,black,dashed]
			(-8,0) arc (270:235:3.314);
			\draw[] (0,-0.2) node {$0$};
			\draw[] (9,-0.2) node {$x$};
			\draw[] (-0.2,7) node {$z$};
			\draw[thin,gray!60]
			(4,0) arc (0:180:6);
			\draw[thin,gray!60]
			(4,0) arc (0:180:4);
			\draw[thin,gray!60]
			(4,0) arc (0:180:2);
			\draw[thin,gray!60]
			(4,0) arc (180:60:4);
			\draw[thin,gray!60]
			(4,0) arc (180:90:6);
			\draw[thin,gray!60]
			(8,0) arc (0:180:6);
			\draw[thin,gray!60]
			(8,0) arc (0:180:4);
			\draw[thin,gray!60]
			(8,0) arc (0:180:2);
			\draw[thin,gray!60]
			(8,0) arc (180:90:2);
			\draw[thin,gray!60]
			(8,0) arc (180:120:4);
			\draw[thin,gray!60]
			(8,0) arc (180:132:6);
			\draw[thin,gray!60]
			(0,0) arc (180:48:6);
			\draw[thin,gray!60]
			(0,0) arc (0:132:6);
			\draw[thin,gray!60]
			(0,0) arc (0:180:4);
			\draw[thin,gray!60]
			(0,0) arc (0:180:2);
			\draw[thin,gray!60]
			(-4,0) arc (0:180:2);
			\draw[thin,gray!60]
			(-4,0) arc (0:120:4);
			\draw[thin,gray!60]
			(-4,0) arc (0:90:6);
			\draw[thin,gray!60]
			(-8,0) arc (0:90:2);
			\draw[thin,gray!60]
			(-8,0) arc (0:60:4);
			\draw[thin,gray!60]
			(-8,0) arc (0:48:6);\draw[thick,->,>=stealth]
			(-10,0)--(10,0);
			\draw[thick,->,>=stealth]
			(0,0)--(0,8);
		\end{tikzpicture}
		\caption{In our new prescription of the holographic Weyl transformed CFT, the PEE network, which is the same as the network in holographic CFT, is represented by gray solid lines. Boundary points are replaced by cutoff spheres (black dashed circles) tangent at the corresponding positions. Under the specific choice of $\phi$ in \eqref{scalar field}, only cutoff spheres at $x<0$ have non-vanishing radius.}
		\label{fig:new prescription}
	\end{figure}

	We give an illustration of our new prescription in Fig.\ref{fig:new prescription}. Now we examine our new prescription and discuss its relations with the basic proposals in \cite{Basu:2023wmv} in the next few subsections.
	
	 \begin{figure}
		\centering
		\begin{tikzpicture}[scale=1.2]
			\draw[thick]
			(-3,0)--(3,0);
			\draw[thick,black,dashed]
			(2.5,0.5) arc (0:360:0.5);
			\draw[thick,black,dashed]
			(-1,1) arc (0:360:1);
			\draw[orange,thick]    (-1.011,1.151) .. controls (-0.3033,-0.045) and (0.1167,1.755) .. (1.54,0.697) ;
			\draw[thick,dashed,orange]
			(-2,0) arc (180:94.366:1.176);
			\draw[thick,dashed,orange]
			(2,0) arc (0:52.843:0.83);
			\draw[thick,blue!60]
			(1.765,0.941) arc (28.08:126.87:2);
			\draw[thick,blue!60,dashed]
			(2,0) arc (0:28.08:2);
			\draw[thick,blue!60,dashed]
			(-2,0) arc (180:126.87:2);
			\filldraw[black] (2,0) circle (1pt);
			\filldraw[orange] (1.54,0.697) circle (1pt);
			\filldraw[orange] (-1.011,1.151) circle (1pt);
			\filldraw[black] (-2,0) circle (1pt);
			\filldraw[blue] (1.765,0.941) circle (1pt);
			\filldraw[blue] (-1.2,1.6) circle (1pt);
			\draw[] (2,-0.2) node {$b$};
			\draw[] (-2,-0.2) node {$a$};
			\draw[] (1,2) node {\textcolor{blue}{$\tilde\Sigma$}};
			\draw[] (-2,1) node {\textcolor{blue}{$\tilde\Sigma_a$}};
			\draw[] (2.2,0.6) node {\textcolor{blue}{$\tilde\Sigma_b$}};
			\draw[] (0.6,1.2) node {\textcolor{orange}{$\Sigma$}};
			\draw[] (-1.3,0.7) node {\textcolor{orange}{$\Sigma_a$}};
			\draw[] (1.7,0.3) node {\textcolor{orange}{$\Sigma_b$}};
		\end{tikzpicture}
		\caption{$\tilde\Sigma$ (blue solid line) is a geodesic whose extended geodesics $\tilde\Sigma_a$ and $\tilde\Sigma_b$ (blue dashed line) end at $x=a$ and $x=b$ respectively. $\Sigma$ (orange solid line) is an arbitrary homologous line anchored on the two cutoff spheres $\bigodot_{a}$ and $\bigodot_{b}$ with $\Sigma_a,\Sigma_b$ (orange dashed line) being the extended geodesics connecting the endpoints of $\Sigma$ with corresponding boundary points. In fact, $\tilde\Sigma$ is still the shortest one among all homologous surfaces connecting the two cutoff spheres.} 
		\label{fig:anchor on cutoff-sphere}
	\end{figure}
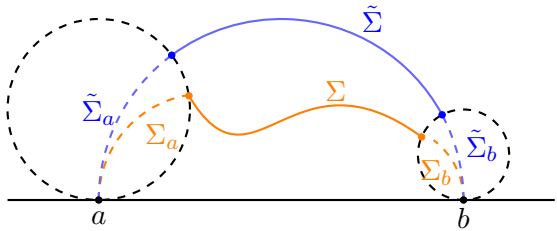
	
	\subsection{Two-point functions and the basic proposal 1}
 Let us first consider the two-point functions of twist operators in the Weyl transformed CFT$_2$, which is given by \eqref{entropy under Weyl transformation}. The geometric picture for the two-point functions was proposed in \cite{Basu:2022crn}. More explicitly, let us consider the geodesic connecting the boundary points $x=a,b$, and decompose it into three parts $\tilde{\Sigma}_a\cup \tilde\Sigma\cup\tilde\Sigma_b$, where the length of $\tilde{\Sigma}_{a,b}$ are $\abs{\phi(a)},\abs{\phi(b)}$ respectively. Then the two-point function $\tilde{S}_{[a,b]}$ \eqref{entropy under Weyl transformation} is reproduced by the length of $\tilde{\Sigma}$ (see also the right figure in Fig.\ref{fig:cutoff-sphere}), i.e.
 \begin{align}\label{holorapictwopoint}
 \tilde{S}_{[a,b]}=&\frac{\text{Area}(\tilde{\Sigma})}{4G}\,.
 \end{align}
 
 In a holographic CFT$_2$, the holographic two-point function of twist operators can be calculated via identifying the saddle point of the gravitational path-integral, and the result is just the minimal surface (or RT surface) homologous to the interval bounded by the two points. This result is equivalent to identifying the homologous surface that intersect with the PEE network for minimal number of times. In the holographic Weyl transformed CFT$_2$, based on our new prescription for the PEE threads it is natural to interpret the two-point function \eqref{entropy under Weyl transformation}  as the homologous surface anchored on the cutoff spheres $\bigodot_{a}$ and $\bigodot_{b}$ that has the minimal number of intersections with the PEE network (or minimal length). 
 
 This indeed coincides with \eqref{holorapictwopoint}. According to our new prescription, since the boundary points are replaced by cutoff spheres, it is natural to consider arbitrary homologous surfaces $\Sigma$ anchored on the two cutoff spheres and optimize the length among all possible $\Sigma$ (see Fig.\ref{fig:anchor on cutoff-sphere}). To compare with the geodesic connecting boundary points, we extend $\Sigma$ to the two boundary points via the two geodesic chords $\Sigma_{a,b}$, whose length equals to $\tilde{\Sigma}_{a,b}$ according to the definition of the cutoff sphere. Since  $\tilde{\Sigma}_a\cup\tilde{\Sigma}\cup  \tilde{\Sigma}_b$ is the minimal surface connecting the two boundary points, the length of $\Sigma_a\cup\Sigma\cup\Sigma_b$ is larger unless $\Sigma$ coincides with $\tilde{\Sigma}$. Therefore, we get to the conclusion that
 \begin{itemize}
 	\item \textit{Between two arbitrary cutoff spheres $\bigodot_{a}$ and $\bigodot_{b}$, the homologous surface that has the minimal length (or the bottle neck) is the portion $\tilde{\Sigma}$ of the geodesic connecting the corresponding two boundary points $x=a,b$.}
 \end{itemize}
Then the two-point function can also be captured by
 \begin{align}
 		\tilde{S}_{[a,b]}=&\min_{\Sigma}\frac{\text{Area}(\Sigma)}{4G}=\frac{\text{Area}(\tilde{\Sigma})}{4G}\,.
 \end{align}
 On the other hand, since the PEE network is not changed, according to \eqref{fluxSigma} the number of intersections between any homologous surface $\Sigma$ and the PEE network is always given by $\text{Area}(\Sigma)/4G$. Hence the two-point function \eqref{entropy under Weyl transformation} is given by the length of the homologous surface that has the minimized number of intersections with the PEE network,
 \begin{align}\label{twopointflux}
 	\tilde{S}_{[a,b]}=&\min_{\Sigma}L(\Sigma)=\min_{\Sigma}\frac{1}{2}\int_{\partial \mathcal{M}}d \textbf{x}\int_{\partial \mathcal{M}} d \textbf{y}\:\omega_{\Sigma}\left(\bf x,\bf y\right)\mathcal{I}\left(\bf x,\bf y\right)
 	\cr
 	=&\frac{\text{Area}(\tilde{\Sigma})}{4G}\,.
 \end{align}
 
\begin{figure}
 	\centering
 	\begin{tikzpicture}[scale=1.2]
 		\draw[thick]
 		(-3,0)--(4,0);
 		\draw[thick,black,dashed]
 		(2.5,0.5) arc (0:360:0.5);
 		\draw[thick,black,dashed]
 		(-1,1) arc (0:360:1);
 		\draw[thick,blue!60]
 		(1.765,0.941) arc (28.08:126.87:2);
 		\draw[thick,blue!60,dashed]
 		(2,0) arc (0:28.08:2);
 		\draw[thick,blue!60,dashed]
 		(-2,0) arc (180:126.87:2);
 		\filldraw[black] (2,0) circle (1pt);
 		\filldraw[black] (-2,0) circle (1pt);
 		\filldraw[blue] (1.765,0.941) circle (1pt);
 		\filldraw[blue] (-1.2,1.6) circle (1pt);
 		\draw[] (2,-0.2) node {$b$};
 		\draw[] (-2,-0.2) node {$a$};
 		\draw[] (1,2) node {\textcolor{blue}{$\tilde\Sigma$}};
 		\draw[] (-2,1) node {\textcolor{blue}{$\tilde\Sigma_a$}};
 		\draw[] (2.2,0.6) node {\textcolor{blue}{$\tilde\Sigma_b$}};
 		\draw[gray]
 		(3,0) arc (0:180:1.5);
 		\draw[gray]
 		(-1.6,0) arc (180:0:2.5);
 		\draw[gray,dashed]
 		(-1.6,0) arc (0:180:0.4);
 		\filldraw[gray] (-0.8,1.82) circle (1pt);
 		\filldraw[gray] (1.35,1.48) circle (1pt);
 	\end{tikzpicture}
 	\caption{The length of the RT surface $\tilde{\Sigma}$ is given by the number of the PEE threads passing through it, for instance the gray solid PEE threads. The gray dashed PEE thread only passes through $\tilde{\Sigma}_a$, but not $\tilde{\Sigma}$ , hence does not contribute to $\tilde{S}_A$. }
 	\label{bottle neck}
 \end{figure}
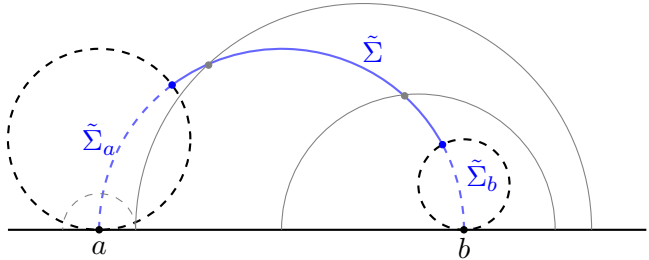

 Based on the PEE thread configuration, we can also directly calculate the PEE $\mathcal{I}(A,\bar{A})$ between an interval $A$ and its complement $\bar{A}$. In the holographic CFT$_2$, this is exactly number of the PEE threads connecting $A$ and $\bar{A}$. In the Weyl transformed case, it is nature to define $\mathcal{I}(A,\bar{A})$ in the following way. 
\begin{itemize}
	\item \textit{Since the endpoints of $A=[a,b]$ are replaced by cutoff spheres, the PEE $\mathcal{I}(A,\bar{A})$ in the Weyl transformed CFT$_2$ is given by the number of the PEE threads connecting $A$ and $\bar{A}$ passing through the bottle neck $\tilde{\Sigma}$, see Fig.\ref{bottle neck}}.
\end{itemize} 
Note that $\tilde{\Sigma}$ is a portion of the geodesic connecting the boundary points of $A$, then according to the discussions in \cite{Lin:2024dho}, the PEE threads intersecting with $\tilde{\Sigma}$ only intersect with $\tilde{\Sigma}$ once. This means the PEE threads connecting $A$ and $\bar{A}$ through $\tilde{\Sigma}$ include all the PEE threads intersecting with $\tilde{\Sigma}$, hence giving the value $\frac{\text{Area}(\tilde{\Sigma})}{4G}$. In other words, we have
\begin{align}\label{twopointI}
\tilde{S}_{[a,b]}=\frac{\text{Area}(\tilde{\Sigma})}{4G}=\mathcal{I}(A,\bar{A})\,.
\end{align} This is exactly the basic proposal 1 \eqref{Basic proposal 1} proposed in \cite{Basu:2023wmv}. When the radius of the cutoff spheres vanishes, $\mathcal{I}(A,\bar{A})$ reproduces the results in holographic CFT$_2$.

Compared with the PEE $\mathcal{I}^{\text{CFT}}(A,\bar{A})$ in CFT$_2$\footnote{Hereafter we use superscript ``CFT'' to distinguish quantities in CFT and quantities in the Weyl transformed CFT.}, the contributions from the PEE threads connecting $\bar{A}$ and $A$ intersecting with $\tilde{\Sigma}_{a,b}$ (see also Fig.\ref{fig:anchor on cutoff-sphere}) are excluded by the Weyl transformation, hence we have
 \begin{equation}\label{I Weyl A Abar 3}
	\mathcal{I}(A,\bar{A})=\mathcal{I}^{\text{CFT}}(A,\bar{A})-\frac{c}{6}|{\phi(a)}|-\frac{c}{6}|{\phi(b)}|\,.
\end{equation}
Note that for a fixed endpoint $x=a$, $\tilde{\Sigma}_{a}$ changes with the choice of the other end-point $x=b$, hence the set of the excluded PEE threads across the end point $x=a$ changes with $b$. In other words, given a boundary point and the corresponding cutoff sphere, we can not determine which classes of PEE threads should be excluded by the Weyl transformation without knowing the other endpoint.  Nevertheless, since the length of $\tilde{\Sigma}_{a}$ is independent from $b$, the number of excluded PEE threads across $x=a$ is fixed and independent from $b$.

\subsection{Partial entanglement entropy between sub-intervals}

\begin{figure}
	\centering
		\begin{tikzpicture}[scale=1.2]
		\draw[thick]
		(-5,0)--(4,0);
		\draw[thick,black,dashed]
		(1.5,0.5) arc (0:360:0.5);
		\draw[thick,black,dashed]
		(3.8,0.8) arc (0:360:0.8);
		\draw[thick,black,dashed]
		(-1.4,0.6) arc (0:360:0.6);
		\draw[thick,black,dashed]
		(-3,1) arc (0:360:1);
		\draw[thick,blue!60,dashed]
		(-4,0) arc (180:136.397:2.5);
		\draw[thick,blue!60]
		(-3.31,1.724) arc (136.397:22.62:2.5);
		\draw[thick,blue!60,dashed]
		(1,0) arc (0:22.62:2.5);
		\draw[thick,blue!60,dashed]
		(-2,0) arc (180:153.009:2.5);
		\draw[thick,blue!60]
		(-1.728,1.135) arc (153.009:35.49:2.5);
		\draw[thick,blue!60,dashed]
		(3,0) arc (0:35.49:2.5);
		\filldraw[black] (1,0) circle (1pt);
		\filldraw[black] (3,0) circle (1pt);
		\filldraw[black] (-2,0) circle (1pt);
		\filldraw[black] (-4,0) circle (1pt);
		\filldraw[blue!60] (0.808,0.962) circle (1pt);
		\filldraw[blue!60] (-3.31,1.724) circle (1pt);
		\filldraw[blue!60] (-1.728,1.135) circle (1pt);
		\filldraw[blue!60] (2.536,1.451) circle (1pt);
		\draw[] (-0.5,-0.25) node {$A_2$};
		\draw[] (2,-0.25) node {$A_3$};
		\draw[] (-3,-0.25) node {$A_1$};
		\draw[] (-4.5,-0.25) node {$\bar{A}$};
		\draw[] (3.7,-0.25) node {$\bar{A}$};
		\draw[] (-4,-0.2) node {$a$};
		\draw[] (-2,-0.2) node {$b$};
		\draw[] (1,-0.2) node {$c$};
		\draw[] (3,-0.2) node {$d$};
		\draw[] (1.8,2.4) node {\textcolor{blue!60}{$\Sigma_2$}};
		\draw[] (-1.8,2.7) node {\textcolor{blue!60}{$\Sigma_1$}};
		\draw[thick,blue!60,dashed]
		(3,0) arc (0:77.32:1);
		\draw[thick,blue!60]
		(2.22,0.98) arc (77.32:126.87:1);
		\draw[thick,blue!60,dashed]
		(1,0) arc (180:126.87:1);
		\filldraw[blue!60] (2.22,0.98) circle (1pt);
		\filldraw[blue!60] (1.4,0.8) circle (1pt);
		\draw[thick,blue!60,dashed]
		(-2,0) arc (0:61.93:1);
		\draw[thick,blue!60]
		(-2.53,0.88) arc (61.93:90:1);
		\draw[thick,blue!60,dashed]
		(-4,0) arc (180:90:1);
		\filldraw[blue!60] (-2.53,0.88) circle (1pt);
		\filldraw[blue!60] (-3,1) circle (1pt);
		\draw[] (-2.7,1.15) node {\textcolor{blue!60}{$\Sigma_3$}};
		\draw[] (2,1.2) node {\textcolor{blue!60}{$\Sigma_4$}};
	\end{tikzpicture}
	\caption{The boundary is decomposed as $A_1\cup A_2\cup A_3\cup \bar{A}$, and the subgions $A_1\cup A_2,A_2\cup A_3,A_2,A_3$ have corresponding RT surfaces $\Sigma_1,\Sigma_2,\Sigma_3,\Sigma_4$ respectively. }
	\label{fig:EL E ER F}
\end{figure}
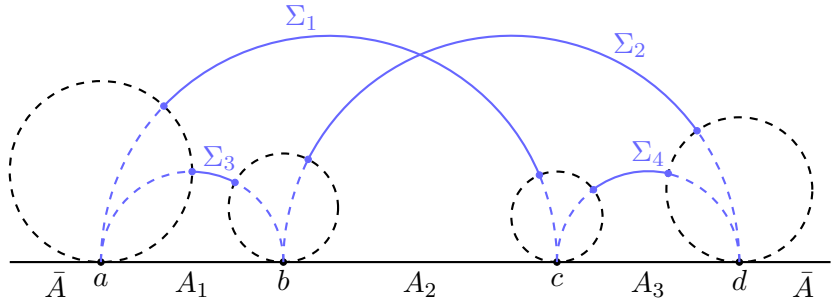

 Now we discuss the PEE between two sub-intervals. Note that, since the Weyl transformation makes the cutoff scale finite, the distance between the endpoints of the sub-intervals under consideration should be bounded from below. A reasonable requirement is that, the two-point functions between any two boundary points of the intervals should be non-negative, i.e. $\tilde{S}_{[a,b]}\geq 0$, for any pair of endpoints $x=a,b$. Also note that, the PEE between disconnected intervals is not the four-point function of twist operators settled at all the boundary endpoints in the Weyl transformed CFT$_2$. We will discuss the multi-point functions in the next subsection.
 
 Let us consider the boundary decomposed into $A_1\cup A_2\cup A_3\cup\bar{A}$ as shown in Fig.\ref{fig:EL E ER F}. In holographic CFT$_2$, the PEE $\mathcal{I}^{\text{CFT}}(A_2,\bar{A})$ is just the number of PEE threads connecting $A_2$ and $\bar{A}$. According to the ALC proposal \eqref{ALC}, it can also be written as a linear combination of subset entanglement entropies (or two-point functions of twist operators). In the Weyl transformed CFT$_2$, the property of additivity indicates that the PEE $\mathcal{I}(A_2,\bar{A})$ can also be written as an additive linear combination of two-point functions, 
 \begin{equation}
 	\begin{aligned}
 		\mathcal{I}(A_2,\bar{A})=&\frac{1}{2}\left(\mathcal{I}(A_1A_2,A_3\bar{A})+\mathcal{I}(A_2A_3,A_1\bar{A})-\mathcal{I}(A_1,A_2A_3\bar{A})-\mathcal{I}(A_3,A_1A_2\bar{A})\right)
 		\\
 		=&\frac{1}{2}\left(\tilde{S}_{A_1A_2}+\tilde{S}_{A_2A_3}-\tilde{S}_{A_1}-\tilde{S}_{A_3}\right)
 		\\
 		=&\frac{1}{2}\left(L(\Sigma_1)+L(\Sigma_2)-L(\Sigma_3)-L(\Sigma_4)\right)\\
 		=&\frac{1}{2}\bigg[\left(\frac{c}{3}\log\frac{d-b}{\epsilon}-\frac{c}{6}|\phi(d)|-\frac{c}{6}|\phi(b)|\right)+\left(\frac{c}{3}\log\frac{c-a}{\epsilon}-\frac{c}{6}|\phi(c)|-\frac{c}{6}|\phi(a)|\right)\\
 		&~~~~-\left(\frac{c}{3}\log\frac{d-c}{\epsilon}-\frac{c}{6}|\phi(d)|-\frac{c}{6}|\phi(c)|\right)-\left(\frac{c}{3}\log\frac{b-a}{\epsilon}-\frac{c}{6}|\phi(b)|-\frac{c}{6}|\phi(a)|\right)\bigg]\\
 		=&\frac{c}{6}\log \frac{(d-b)(c-a)}{(d-c)(b-a)}\,,
 	\end{aligned}
 \end{equation}
 where we have used the basic proposal 1. We find that all the Weyl scalar terms cancel with each other, hence we reproduce exactly the same PEE as in CFT$_2$
 \begin{align}
 	 		\mathcal{I}(A_2,\bar{A})=&\mathcal{I}^{\text{CFT}}(A_2,\bar{A})\,.
 \end{align}
 This result implies that the PEE between intervals that has the distance larger than the cutoff scale is not affected by the Weyl transformation. This is reasonable as the PEE between disconnected intervals are correlation at the scale larger than the cutoff scale, while the Weyl transformation only affect the correlation at scale smaller than the cutoff scale.

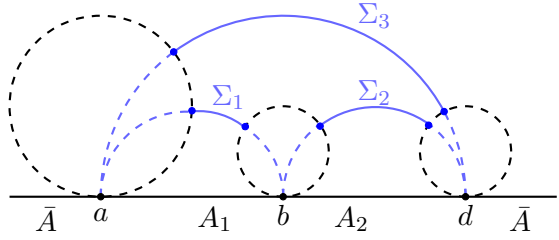
\begin{figure}
  	\centering
  	\begin{tikzpicture}[scale=1.2]
  		\draw[thick]
  		(-3,0)--(3,0);
  		\draw[thick,black,dashed]
  		(2.5,0.5) arc (0:360:0.5);
  		\draw[thick,black,dashed]
  		(0.5,0.5) arc (0:360:0.5);
  		\draw[thick,black,dashed]
  		(-1,1) arc (0:360:1);
  		\draw[thick,blue!60]
  		(-0.4,0.75) arc (49:95:0.8);
  		\draw[thick,blue!60,dashed]
  		(0,0) arc (0:60:0.9);
  		\draw[thick,blue!60,dashed]
  		(-2,0) arc (180:85:0.94);
  		\draw[thick,blue!60]
  		(1.765,0.941) arc (28.08:126.87:2);
  		\draw[thick,blue!60,dashed]
  		(2,0) arc (0:28.08:2);
  		\draw[thick,blue!60,dashed]
  		(-2,0) arc (180:126.87:2);
  		\draw[thick,blue!60]
  		(1.6,0.8) arc (54:126:1);
  		\draw[thick,blue!60,dashed]
  		(0,0) arc (180:126.87:1);
  		\draw[thick,blue!60,dashed]
  		(2,0) arc (0:60:1);
  		\filldraw[black] (2,0) circle (1pt);
  		\filldraw[black] (0,0) circle (1pt);
  		\filldraw[black] (-2,0) circle (1pt);
  		\filldraw[blue] (1.765,0.941) circle (1pt);
  		\filldraw[blue] (-1.2,1.6) circle (1pt);
  		\filldraw[blue] (1.595,0.79) circle (1pt);
  		\filldraw[blue] (0.405,0.78) circle (1pt);
  		\filldraw[blue] (-1,0.95) circle (1pt);
  		\filldraw[blue] (-0.415,0.775) circle (1pt);
  		\draw[] (0,-0.2) node {$b$};
  		\draw[] (2,-0.2) node {$d$};
  		\draw[] (-2,-0.2) node {$a$};
  		\draw[] (1,2) node {\textcolor{blue!60}{$\Sigma_3$}};
  		\draw[] (-0.6,1.1) node {\textcolor{blue!60}{$\Sigma_1$}};
  		\draw[] (1,1.15) node {\textcolor{blue!60}{$\Sigma_2$}};
  		\draw[] (-0.75,-0.25) node {$A_1$};
  		\draw[] (0.75,-0.25) node {$A_2$};
  		\draw[] (2.6,-0.25) node {$\bar{A}$};
  		\draw[] (-2.6,-0.25) node {$\bar{A}$};
  	\end{tikzpicture}
  	\caption{The boundary is decomposed as $A_1\cup A_2\cup \bar{A}$ with corresponding RT surfaces $\Sigma_1,\Sigma_2,\Sigma_3$ respectively. }
  	\label{fig:interval EF}
  \end{figure}

Similarly, when the two subregions are adjacent as in Fig.\ref{fig:interval EF}, one can also apply the additivity property to compute $\mathcal{I}(A_1,\bar{A})$ via two-point functions of twist operators, i.e. 
 \begin{equation}
 	\begin{aligned}
 		\mathcal{I}(A_1,\bar{A})=&\frac{1}{2}\left(\mathcal{I}(A_1A_2,\bar{A})+\mathcal{I}(A_1,A_2\bar{A})-\mathcal{I}(A_2,A_1\bar{A})\right)
 		\\
 		=&\frac{1}{2}\left(\tilde{S}_{A_1A_2}+\tilde{S}_{A_1}-\tilde{S}_{A_2}\right)
 		\\
 		=&\frac{c}{6}\log \frac{(d-a)(b-a)}{(d-b)\epsilon}-\frac{c}{6}|\phi(a)|
 		\\
 		=&\mathcal{I}^{\text{CFT}}(A_1,\bar{A})-\frac{c}{6}|\phi(a)|\,.
 	\end{aligned}
 \end{equation} 
One can also calculate $\mathcal{I}(A_2,\bar{A})$ in the same way and get
 \begin{equation}
	\begin{aligned}
		\mathcal{I}(A_2,\bar{A})=&\frac{1}{2}\left(\mathcal{I}(A_1A_2,\bar{A})+\mathcal{I}(A_2,A_1\bar{A})-\mathcal{I}(A_1,A_2\bar{A})\right)
		\\
		=&\frac{1}{2}\left(\tilde{S}_{A_1A_2}+\tilde{S}_{A_2}-\tilde{S}_{A_1}\right)
		\\
		=&\frac{c}{6}\log \frac{(d-a)(d-b)}{(b-a)\epsilon}-\frac{c}{6}|\phi(d)|
		\\
		=&\mathcal{I}^{\text{CFT}}(A_2,\bar{A})-\frac{c}{6}|\phi(d)|\,.
	\end{aligned}
\end{equation} 
The summation of these two PEEs is given by,
 \begin{equation}
 	\begin{aligned}
 		&\mathcal{I}(A_1,\bar{A})+\mathcal{I}(A_2,\bar{A})
 		\\=&\mathcal{I}^{\text{CFT}}(A_1,\bar{A})+\mathcal{I}^{\text{CFT}}(A_2,\bar{A})-\frac{c}{6}|\phi(a)|-\frac{c}{6}|\phi(d)|
 		\\
 		=& \mathcal{I}^{\text{CFT}}(A,\bar{A})-\frac{c}{6}|\phi(a)|-\frac{c}{6}|\phi(d)|
 	\end{aligned}
 \end{equation}
  which reproduces our previous result \eqref{I Weyl A Abar 3} for $\mathcal{I}(A,\bar{A})$ and is consistent with the property of additivity.
  
  To summarize, each connecting point between subregions contributes to a regulation term as $\frac{c}{6}\phi$ such that a generic PEE between two arbitrary single intervals $E,F$ is regulated as follows,
 \begin{equation}\label{regulatedPEE}
	\mathcal{I}(E,F)=\mathcal{I}^{\text{CFT}}(E,F)-\begin{cases}\left(\frac{c}{6}{|\phi(a)|}+\frac{c}{6}{|\phi(b)|}\right), & \text{if } E\cap F=\{x=a,x=b\}\\ \frac{c}{6}{|\phi(a)|}, & \text{if }E\cap F=\{x=a\} \\ 0, & \text{if }E\cap F=\emptyset \end{cases}
 \end{equation}
 which describes the PEE when: 1) $F=\bar{E}$; 2) $E$ and $F$ are adjacent intervals; 3) $E$ and $F$ are non-adjacent intervals. If $E$ (or $F$) is a multi-interval, the PEE can be constructed after decomposition into several PEEs between single intervals due to the property of additivity.
 
   \subsection{Four-point functions and the basic proposal 2}
   
   Now we turn to the four-point function $\tilde{S}_{[d,c]\cup[a,b]}$ for twist operators. In the holographic CFT$_2$, this four-point function corresponds to the holographic entanglement entropy for two disconnected intervals, which undergoes a phase transition between disconnected and connected entanglement wedge,
   \begin{align}\label{four-point-f}
   	\tilde{S}_{[d,c]\cup[a,b]} &=\min \{\tilde{S}_{[d,c]}+\tilde{S}_{[a,b]},\tilde{S}_{[d,b]}+\tilde{S}_{[c,a]}\}\,.
   \end{align}
   Note that in this case the two-point function of any single interval is just the entanglement entropy, i.e. $\tilde{S}_{A}=S_{A}$. The above two phases correspond to two different extreme surfaces homologous to boundary multi-intervals. This simple result can be obtained in the CFT side under the large $c$ limit \cite{Hartman:2013mia,Faulkner:2013yia}. 
   
   In the holographic Weyl transformed CFT$_2$, we should also take the large $c$ limit and it is reasonable to assume that, the four-point function can be decomposed into a similar linear combination of two-point functions as \eqref{four-point-f}. Based on our new prescription, the two-point functions in \eqref{four-point-f} are given by the area of the minimal homologous surfaces anchored on the corresponding cutoff spheres, and the four-point function can be interpreted as the minimal number of intersections among all possible surfaces homologous to the two-interval. This prescription can be extended to multi-point correlation functions, which is also an optimization of all possible homologous surfaces, as an extension of the RT formula for multi-intervals.
   
    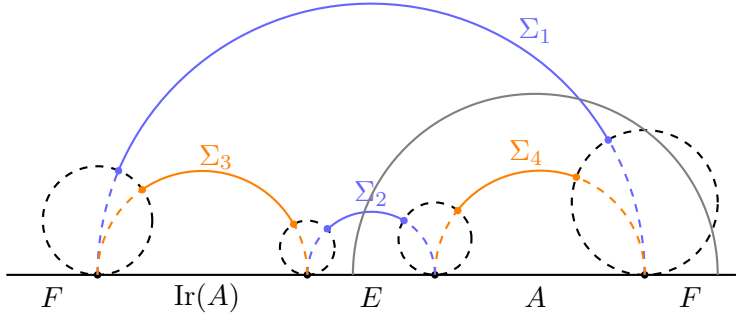
\begin{figure}
   	\centering
   	\begin{tikzpicture}[scale=1.2]
   		\draw[thick]
   		(-4,0)--(4,0);
   		\draw[thick,black,dashed]
   		(3.8,0.8) arc (0:360:0.8);
   		\draw[thick,black,dashed]
   		(-2.4,0.6) arc (0:360:0.6);
   		\draw[thick,black,dashed]
   		(1.1,0.4) arc (0:360:0.4);
   		\draw[thick,black,dashed]
   		(-0.4,0.3) arc (0:360:0.3);
   		\filldraw[black] (0.7,0) circle (1pt);
   		\filldraw[black] (3,0) circle (1pt);
   		\filldraw[black] (-0.7,0) circle (1pt);
   		\filldraw[black] (-3,0) circle (1pt);
   		\draw[thick,blue!60,dashed]
   		(3,0) arc (0:29.86:3);
   		\draw[thick,blue!60]
   		(2.6,1.49) arc (29.86:157.38:3);
   		\draw[thick,blue!60,dashed]
   		(-3,0) arc (180:157.38:3);
   		\filldraw[blue!60] (2.6,1.49) circle (1pt);
   		\filldraw[blue!60] (-2.77,1.15) circle (1pt);
   		\draw[thick,orange,dashed]
   		(3,0) arc (0:69.65:1.15);
   		\draw[thick,orange]
   		(2.25,1.08) arc (69.65:141.64:1.15);
   		\draw[thick,orange,dashed]
   		(0.7,0) arc (180:141.64:1.15);
   		\filldraw[orange] (2.25,1.08) circle (1pt);
   		\filldraw[orange] (0.95,0.71) circle (1pt);
   		\draw[thick,blue!60,dashed]
   		(0.7,0) arc (0:59.49:0.7);
   		\draw[thick,blue!60]
   		(0.36,0.6) arc (59.49:133.6:0.7);
   		\draw[thick,blue!60,dashed]
   		(-0.7,0) arc (180:133.6:0.7);
   		\filldraw[blue!60] (0.36,0.6) circle (1pt);
   		\filldraw[blue!60] (-0.48,0.51) circle (1pt);
   		\draw[thick,orange,dashed]
   		(-0.7,0) arc (0:29.24:1.15);
   		\draw[thick,orange]
   		(-0.85,0.56) arc (29.24:124.9:1.15);
   		\draw[thick,orange,dashed]
   		(-3,0) arc (180:124.9:1.15);
   		\filldraw[orange] (-2.51,0.94) circle (1pt);
   		\filldraw[orange] (-0.85,0.56) circle (1pt);
   		\draw[] (0,-0.25) node {$E$};
   		\draw[] (1.8,-0.25) node {$A$};
   		\draw[] (-1.8,-0.25) node {Ir$(A)$};
   		\draw[] (3.5,-0.25) node {$F$};
   		\draw[] (-3.5,-0.25) node {$F$};
   		\draw[] (1.8,2.7) node {\textcolor{blue!60}{$\Sigma_1$}};
   		\draw[] (0,0.9) node {\textcolor{blue!60}{$\Sigma_2$}};
   		\draw[] (1.7,1.35) node {\textcolor{orange}{$\Sigma_4$}};
   		\draw[] (-1.7,1.3) node {\textcolor{orange}{$\Sigma_3$}};
   		\draw[thick,gray]
   		(3.8,0) arc (0:180:2);
   	\end{tikzpicture}
   	\caption{There are two possible extremal surfaces $\Sigma_1\cup\Sigma_2$ (blue solid lines) and $\Sigma_3\cup\Sigma_4$ (orange solid lines), which are both homologous to $A\cup\text{Ir}\left(A\right)$. Here $\Sigma_1\cup\Sigma_2$ is the RT surface whose length is always smaller than that for $\Sigma_3\cup\Sigma_4$. There are some PEE threads connecting $E$ with $F$ passing through $\Sigma_1\cup\Sigma_2$, for instance the gray solid PEE thread, hence contributing to $\tilde{S}_{A\text{Ir}\left(A\right)}$.}
   	\label{A and its island}
   \end{figure}
   
   For the specific configurations considered in \cite{Basu:2023wmv}, the region $[d,c]$ is the generalized island region Ir$(A)$ for the interval $A=[a,b]$, which means $A$ admits an island region. In such configurations the connected phase always gives the smaller four-point function, see Fig.\ref{A and its island}, hence
   \begin{align}
   	\tilde{S}_{A\cup\text{Ir}(A)}=\tilde{S}_{[d,b]}+\tilde{S}_{[c,a]}=L(\Sigma_1)+L(\Sigma_2)\,,
   \end{align}
   which is exactly the basic proposal 2 \eqref{Basic proposal 2} proposed in \cite{Basu:2023wmv}. Using the basic proposal 1 \eqref{Basic proposal 1}, we can further relate the four-point function to PEEs,
   \begin{equation}\label{fourpointpee}
   		\begin{aligned}
   			\tilde{S}_{A\cup\text{Ir}(A)}=&\tilde{S}_{[d,b]}+\tilde{S}_{[c,a]}
   	\cr
   	=&\mathcal{I}(\text{Ir}(A)EA,F)+\mathcal{I}(\text{Ir}(A)FA,E)\cr
   	=&\mathcal{I}(\text{Ir}(A)A,EF)+2\mathcal{I}(E,F)\,,
   	\end{aligned}
   \end{equation}   
   where $\mathcal{I}(E,F)$ can be calculated by \eqref{regulatedPEE}. This confirms that, as in the two-interval configuration in holographic CFT$_2$, the normalization property does not apply for disconnected intervals $A\cup\text{Ir}(A)$, and
   \begin{align}
   	\tilde{S}_{A\cup\text{Ir}(A)}>\mathcal{I}(\text{Ir}(A)A,\overline{\text{Ir}(A)A})\,.
   \end{align}
   Also the entanglement contribution is not equivalent to the two-body representation.

   	\section{Island formula in the Weyl transformed CFT}\label{sec:PEE island}

	In this section, we reproduce the island formula for the Weyl transformed CFT$_2$ based on our new prescription. More explicitly, given an region $A$ which can either be a single interval or multi-intervals, and the associated cutoff spheres, we should consider all possible homologous surfaces $\Sigma$. Note that according to our new prescription, the homologous surfaces $\Sigma$ are possible to anchor on the cutoff spheres associated to any boundary points which may not belong to the boundary of $A$. Among all possible $\Sigma$, we optimize the number of intersections between $\Sigma$ and the bulk PEE threads, which is just the length of $\Sigma$. Then the $\tilde{\Sigma}$ with the minimal $L(\Sigma)$ gives the entanglement entropy for $A$, i.e.
	\begin{equation}\label{entanglement entropy under minimal}
		S_A=\min_{\Sigma \sim A} L\left(\Sigma\right)=L(\tilde{\Sigma})=\frac{Area(\tilde{\Sigma})}{4G}\,.
	\end{equation} 
	We will check that, the above prescription will reproduce the holographic picture of island formula in KR braneworld or AdS/BCFT.

	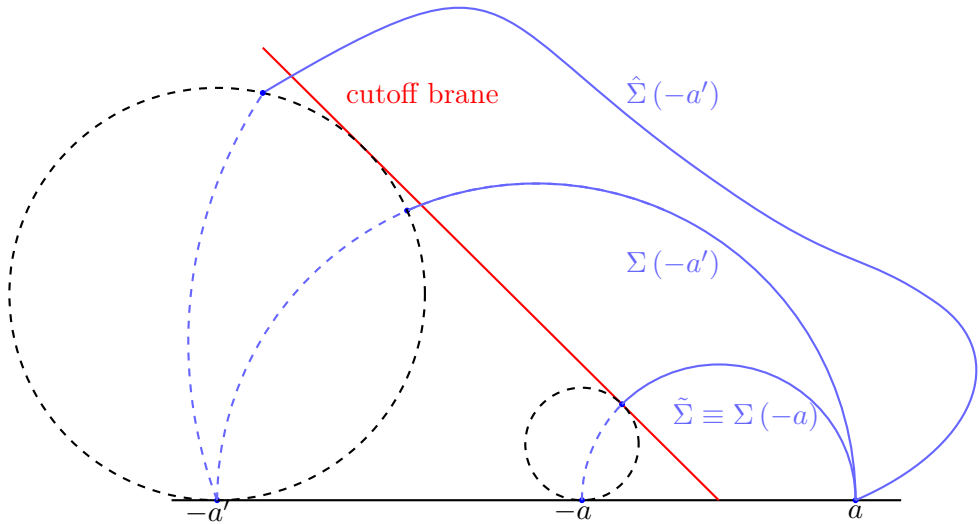
\begin{figure}
		\centering
		\begin{tikzpicture}[scale=0.6]
			\draw[thick]
			(-12,0)--(4,0);
			\draw[thick,red]
			(0,0)--(-10,10);
			\draw[] (-6.5,9) node {\textcolor{red}{cutoff brane}};
			\filldraw[blue] (3,0) circle (1.5pt);
			\filldraw[blue] (-6.834,6.401) circle (1.5pt);
			\filldraw[blue] (-2.121,2.121) circle (1.5pt);
			\filldraw[blue] (-3,0) circle (1.5pt);
			\filldraw[blue] (-11,0) circle (1.5pt);
			\draw[] (-3.2,-0.25) node {$-a$};
			\draw[] (3,-0.25) node {$a$};
			\draw[] (-11.2,-0.25) node {$-a'$};
			\draw[] (0.6,1.9) node {\textcolor{blue!60}{$\tilde{\Sigma}\equiv\Sigma\left(-a\right)$}};
			\draw[] (-1,5.2) node {\textcolor{blue!60}{$\Sigma\left(-a'\right)$}};
			\draw[] (-1,9) node {\textcolor{blue!60}{$\hat{\Sigma}\left(-a'\right)$}};
			\draw[thick,blue!60]
			(3,0) arc (0:135:3);
			\draw[thick,blue!60,dashed]
			(-3,0) arc (180:135:3);
			\draw[thick,blue!60]
			(3,0) arc (0:113.881:7);
			\draw[thick,dashed,blue!60]
			(-11,0) arc (180:66.119:7);
			\draw[thick,dashed]
			(-1.757,1.243) arc (0:360:1.243);
			\draw[thick,dashed]
			(-6.444,4.556) arc (0:360:4.556);
			\draw[yscale=-1,xscale=1,thick,blue!60]    (-10,-9) .. controls (-4.1379,-12.511) and (-5.4367,-10.7116) .. (0.1433,-6.8911) .. controls (2.7234,-5.0706) and (3.1433,-5.5178) .. (4.7967,-4.3444) .. controls (6.45,-3.1711) and (5.6133,-1.1338) .. (3,0) ;
			\draw[thick,dashed,blue!60]
			(-11,0) arc (200.5:147:10);
			\filldraw[blue] (-10,9) circle (1.5pt);
		\end{tikzpicture}
		\caption{$\hat{\Sigma}\left(-a'\right)$ is an arbitrary homologous surface while $\Sigma\left(-a'\right)$ is a geodesic chord, both of which have extended geodesics inside $\bigodot_{-a'}$ end at the boundary point $x=-a'$. The length of $\Sigma\left(-a'\right)$ is strictly smaller than the length of $\hat{\Sigma}\left(-a'\right)$ unless they coincides with each other according to arguments in Fig.\ref{fig:anchor on cutoff-sphere}. $L\left(\Sigma\left(-a'\right)\right)$ with $a'$ being a variable is minimal when $\Sigma\left(-a'\right)$ becomes the RT surface $\tilde{\Sigma}=\Sigma\left(-a\right)$. }
		\label{fig:reformulation}
	\end{figure}
	
	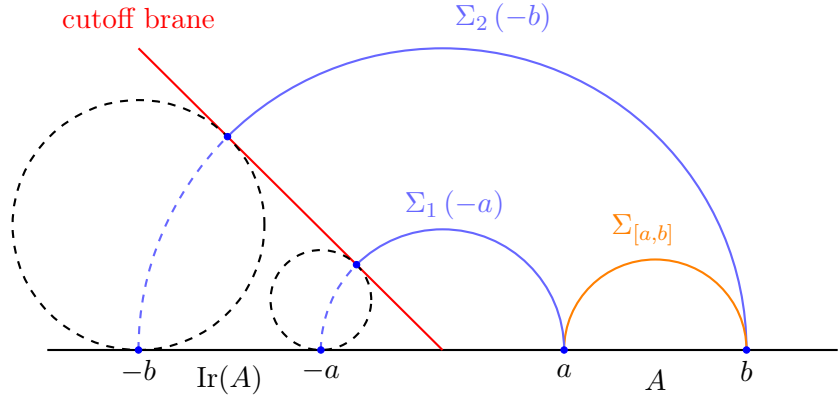
\begin{figure}
		\centering
		\begin{tikzpicture}[scale=0.8]
			\draw[thick]
			(-6.5,0)--(6.5,0);
			\draw[thick,red]
			(0,0)--(-5,5);
			\draw[] (-5,5.5) node {\textcolor{red}{cutoff brane}};
			\draw[thick,blue!60]
			(5,0) arc (0:135:5);
			\draw[thick,orange]
			(5,0) arc (0:180:1.5);
			\draw[thick,dashed,blue!60]
			(-5,0) arc (180:135:5);
			\draw[thick,dashed]
			(-2.929,2.071) arc (0:360:2.071);
			\draw[thick,dashed]
			(-2,0.828) circle (0.828);
			\draw[thick,blue!60]
			(2,0) arc (0:135:2);
			\draw[thick,dashed,blue!60]
			(-2,0) arc (180:135:2);
			\filldraw[blue] (2,0) circle (1.5pt);
			\filldraw[blue] (5,0) circle (1.5pt);
			\filldraw[blue] (-2,0) circle (1.5pt);
			\filldraw[blue] (-5,0) circle (1.5pt);
			\filldraw[blue] (-1.414,1.414) circle (1.5pt);
			\filldraw[blue] (-3.536,3.536) circle (1.5pt);
			\draw[] (2,-0.35) node {$a$};
			\draw[] (3.5,-0.5) node {$A$};
			\draw[] (5,-0.35) node {$b$};
			\draw[] (-3.5,-0.5) node {Ir$\left(A\right)$};
			\draw[] (-2,-0.35) node {$-a$};
			\draw[] (-5,-0.35) node {$-b$};
			\draw[] (1,5.5) node {\textcolor{blue!60}{$\Sigma_{2}\left(-b\right)$}};
			\draw[] (0.2,2.4) node {\textcolor{blue!60}{$\Sigma_{1}\left(-a\right)$}};
			\draw[] (3.3,2.0) node {\textcolor{orange}{$\Sigma_{[a,b]}$}};
		\end{tikzpicture}
		\caption{For a finite interval $A=[a,b]$, the entanglement entropy $S_A$ is given by minimization between two possible phases, namely $\Sigma_{1}\left(-a\right)\cup\Sigma_{2}\left(-b\right)$ and $\Sigma_{[a,b]}$, which coincides with the island formula.}
		\label{fig:reformulation2}
	\end{figure}

Let us consider the Weyl transformation that characterized by the scalar \eqref{scalar field}, and the entanglement entropy for the region $A=\left[a,\infty\right)$ with $a>0$ as shown in Fig.\ref{fig:reformulation}. The minimization procedure of $L\left(\Sigma\right)$ among all possible homologous surfaces can be split into two steps, we first fix $a'$ and find out which homologous surface $\hat{\Sigma}\left(-a'\right)$ anchored on the cutoff spheres $\bigodot_{-a'}$ and $\bigodot_{a}$ \footnote{ Note that, the cutoff sphere $\bigodot_{a}$ is just the boundary point $x=a$ (i.e. cutoff sphere with zero radius).} gives the local minimal length divided by $4G$. According to the discussion of above subsections (arguments around Fig.\ref{fig:anchor on cutoff-sphere}), the geodesic chord $\Sigma\left(-a'\right)$ has the smallest length among all whose extended geodesic ends at the boundary point $x=a'$. Secondly, regard $a'$ as a variable and find out which $a'$ gives the minimal $L\left(\Sigma\left(a'\right)\right)$, see Fig.\ref{fig:reformulation} for an illustration,
		\begin{align}
			S_A=&\min_{a'}\frac{L\left(\Sigma\left(-a'\right)\right)}{4G}=\min_{a'}\frac{\text{Length}\left(\Sigma\left(-a'\right)\right)}{4G}\notag\\
			=&\min_{a'}\left\{\frac{1}{4G}\left(2\log\frac{a+a'}{\epsilon}-\log\frac{2a'}{\epsilon}+\kappa\right)\right\}
			\notag\\
			=&\frac{c}{6}\log\frac{2a}{\epsilon}+\frac{c}{6}\kappa\,,
		\end{align}
		where we have used the solution $a'=a$ to the optimization and $c=\frac{3}{2G}$. One can easily see the expression of above equation is identical to \eqref{shexian}.
		
Next we consider a finite interval $A=\left[a,b\right]$ where $b>a>0$ as shown in Fig.\ref{fig:reformulation2}, and a possible island region $\text{Is}(A)=[-b',-a']$ where $b'>a'>0$. Again we first take a pair of parameters $(a',b')$ and find out the homologous surface $\Sigma_1\left(-a'\right)$ anchored on the $\bigodot_{-a'}$ and $\bigodot_{a}$, and has the  minimal length. According to our previous discussions, $ \Sigma_1\left(-a'\right)$ is just a portion of the geodesic connecting the boundary points $x=-a'$ and $x=a$, which is cut off at $\bigodot_{-a'}$ and $\bigodot_{a}$. Similarly we determine the minimal homologous surface $\Sigma_2\left(-b'\right)$ anchored on $\bigodot_{-b'}$ and $\bigodot_{b}$. Secondly we consider all possible values of $a'$ and $b'$ and find the minimal values for the summation of their length. Then the entanglement entropy $S_A$ should be given by
\begin{equation}
	S_A=\text{min}\left\{\min_{a',b'} \frac{\text{Length} \left(\Sigma_1\left(-a'\right)\cup\Sigma_2\left(-b'\right)\right)	}{4G}, \frac{\text{Length}\left(\Sigma_{[a,b]}\right)}{4G} \right\}\,,
\end{equation}
where $\Sigma_{[a,b]}$ is the minimal homologous surface without island. Since 
\begin{equation}
	\begin{aligned}
		\min_{a',b'}\frac{\text{Length}\left(\Sigma_1\left(-a'\right)\cup\Sigma_2\left(-b'\right)\right)}{4G} 
		&=\min_{a',b'}\frac{c}{6}\left(2\log\frac{a+a'}{\epsilon}-\log\frac{2a'}{\epsilon}+2\log\frac{b+b'}{\epsilon}-\log\frac{2b'}{\epsilon}\right)+\frac{c}{3}\kappa\,\\
		&=\frac{c}{6}\log\frac{2a}{\epsilon}+\frac{c}{6}\log\frac{2b}{\epsilon}+\frac{c}{3}\kappa\\
		&=\frac{\text{Length}\left(\Sigma_1\left(-a\right)\cup\Sigma_2\left(-b\right)\right)}{4G} ,
	\end{aligned}
\end{equation}
where the minimal value of this summations is given by $a'=a$ and $b'=b$ as shown in Fig.\ref{fig:reformulation2}, such that
\begin{align}
	S_A=\text{min}\left\{\frac{c}{6}\log\frac{2a}{\epsilon}+\frac{c}{6}\log\frac{2b}{\epsilon}+\frac{c}{3}\kappa,~ \frac{c}{3}\log \frac{b-a}{\epsilon} \right\}\,,
\end{align}
which gives exactly the island formula in the finite interval configuration.

\section{Entanglement contribution and BPE in island phase revisited}\label{sec.revisited}

\subsection{A brief introduction to BPE in island phase}
Let us briefly introduce the steps to compute the BPE in island phases in \cite{Basu:2023wmv}. We will focus on the Weyl transformed CFT$_2$ with the Weyl scalar \eqref{scalar field}, which simulates the KR braneworld or AdS/BCFT correspondence with a KR brane settled at $\rho=\kappa$. Also we only discuss the case of two adjacent intervals $A\cup B$ on the AdS boundary and $AB$ admits an entanglement island region $\text{Is}(AB)$ (see \cite{Basu:2023wmv} for non-adjacent configurations). We will first introduce the BPE for two adjacent intervals $AB$ in holographic CFT$_2$, then we extend this concept to the island phases in the Weyl transformed case.

We consider the boundary state to be the vacuum of the holographic CFT$_2$ and the complement of $AB$ is divided into $A_1\cup B_1$. It was proposed in \cite{Wen:2021qgx} that, the BPE$(A,B)$ between $A$ and $B$ is a measure of mixed state correlations that holographically duals to the entanglement wedge cross section (EWCS) in the bulk. The BPE$(A,B)$ is given by $s_{AA_1}(A)$ that satisfies the so-called balance requirement:
\begin{equation}\label{bc1}
	s_{AA_1}(A)=s_{BB_1}(B),\quad \text{or}\quad 
	\mathcal{I}(A,BB_1)=\mathcal{I}(B,AA_1).
\end{equation}
This balance requirement is sufficient to determine the partition point in the purifying system $A_1B_1$, which we call the balance point. Note that, in holographic CFT$_2$ the above two equations are equivalent as the two representations for PEE is equivalent. The BPE is then defined as 
\begin{align}\label{BPEd}
	\text{BPE}(A,B)=s_{AA_1}(A)|_{\rm balanced}=s_{BB_1}(B)|_{\rm balanced}
\end{align}

From the viewpoint of quantum information  \cite{Basu:2022crn}, the state in island phase is self-encoded, which means the degrees of freedom in the island region $\is(A)$ is not independent as the state of $\is(A)$ is determined by the state of $A$. It was pointed out in \cite{Basu:2023wmv} that, when we compute the contribution $s_{A}(\alpha)$, we should also take the contribution from the generalized island $\text{Ir} (\alpha)$ into account. Let us start with the region $AB$ and the island regions Is$(AB)$, Is$(A)$ and Is$(B)$. The complement $A_1B_1$ may also admit an island region $\is(A_1B_1)$. The PEE, in contribution representation for example, should contain the contribution from the island region. Nevertheless, there are regions which are included in $\is (AB)$ but outside $\is(A)\cup\is(B)$, which should also play a non-trivial role to the entanglement entropy. In other words, when $\is (AB)\neq \emptyset$ and $\is (AB)\supset \is (A)\cup \is (B)$, we define the ownerless island region to be
\begin{align}
	\text{Io}(AB) =\is(AB)/(\is (A)\cup \is (B))\,.
\end{align}
The ownerless island should be further divided into two parts $\text{Io}(AB)=\text{Io}(A)\cup \text{Io}(B)\,,$ which are assigned to $A$ and $B$ respectively. To conclude, we define the  \emph{generalized islands} (see also the so-called reflected entropy islands defined in \cite{Chandrasekaran:2020qtn})
\begin{align}\label{genisland}
	\mathrm{Ir}(A)=\is(A)\cup \text{Io}(A),\qquad \mathrm{Ir}(B)=\is(B)\cup \text{Io}(B)\,,
\end{align}
as the island region assigned to $A$ and $B$ respectively when calculating the contributions. One can also define the generalized island regions for $A_1$ and $B_1$. The next step is to clarify the rules to identify the generalized (or ownerless) island regions. 

Now we generalize the balance requirements \eqref{bc1} to the island phases, which involves the calculation of the entanglement contribution. For convenience we denote $C \equiv\overline{AB\cup\text{Is}(AB)}$ and consider the contribution $s_{AB}(A)$. After taking into account the contributions from the islands, in \cite{Basu:2023wmv} the authors conducted the following computation,
\begin{equation}\label{galc1}
	\begin{aligned}
		s_{AB}(A)
		=& \mathcal{I}(A\text{Ir}(A),C)\\
		=& \frac{1}{2}\left[
		\mathcal I(A\text{Ir}(A) B \text{Ir}(B), C) + \mathcal I(A\text{Ir}(A), B \text{Ir}(B) C) - \mathcal I(B\text{Ir}(B),A\text{Ir}(A) C)\right]\\
		=& \frac{1}{2}\left[ \tilde S_{A\text{Ir}(A)B\text{Ir}(B)}+\tilde S_{A\text{Ir}(A)}- \tilde S_{B\text{Ir}(B)}\right]\,.
	\end{aligned}
\end{equation}
We call this formula the \emph{generalized ALC (GALC) formula} \cite{Basu:2023wmv} for island phases, which is just the ALC formula \eqref{ALC} with the replacement $S_{\gamma}\Rightarrow \tilde S_{\gamma\text{Ir}(\gamma)}$ applied to each term\footnote{Compared with \eqref{ALC}, here $\alpha_L=\emptyset$, $\alpha_R=B$ which is replaced by $B\text{Ir}(B)$, $\alpha=A$ which is replaced by $A\text{Ir}(A)$.}. 

\begin{figure}
	\centering
	\begin{tikzpicture}[scale=0.8]
		\draw[thick]
		(-8.5,0)--(7.5,0);
		\draw[thick,red]
		(0,0)--(-5.5,5.5);
		\draw[] (-5.5,5.7) node {\textcolor{red}{cutoff brane}};
		\draw[thick,dashed] (-1,0.41) circle (0.41);
		\draw[thick,dashed] (-5.5,2.28) circle (2.28);
		\draw[thick,blue!60]
		(1,0) arc (0:135:1);
		\draw[thick,blue!60,dashed]
		(-1,0) arc (180:135:1);
		\draw[thick,blue!60]
		(5.5,0) arc (0:135:5.5);
		\draw[thick,blue!60,dashed]
		(-5.5,0) arc (180:135:5.5);
		\draw[thick,orange]
		(3.25,0) arc (0:135:3.25);
		\draw[thick,orange,dashed]
		(-3.25,0) arc (180:135:3.25);
		\filldraw[black] (0,0) circle (1pt);
		\filldraw[blue!60] (-0.71,0.71) circle (1pt);
		\filldraw[blue!60] (-3.89,3.89) circle (1pt);
		\filldraw[orange] (-2.3,2.3) circle (1pt);
		\filldraw[black] (1,0) circle (1pt);
		\filldraw[black] (3.25,0) circle (1pt);
		\filldraw[black] (5.5,0) circle (1pt);
		\filldraw[black] (-1,0) circle (1pt);
		\filldraw[black] (-3.25,0) circle (1pt);
		\filldraw[black] (-5.5,0) circle (1pt);
		
		\draw[] (0.55,-0.3) node {\small{$A_1$}};
		\draw[] (2.25,-0.3) node {\small{$A$}};
		\draw[] (4.25,-0.3) node {\small{$B$}};
		\draw[] (6.6,-0.3) node {\small{$B_1$}};
		\draw[] (-0.55,-0.3) node {\small{\text{Ir}$(A_1)$}};
		\draw[] (-2.25,-0.3) node {\small{Ir$(A)$}};
		\draw[] (-4.25,-0.3) node {\small{Ir$(B)$}};
		\draw[] (-6.9,-0.3) node {\small{Ir$(B_1)$}};
		\draw[] (-5.55,0.25) node {\small{$-c$}};
		\draw[] (-3.7,0.25) node {\small{$-q_1$}};
		\draw[] (-1,0.25) node {\small{$-a$}};
		\draw[] (1.2,0.25) node {\small{$a$}};
		\draw[] (3.35,0.25) node {\small{$b$}};
		\draw[] (5.7,0.25) node {\small{$c$}};
		\draw[] (0.8,1) node {\textcolor{blue!60}{$\Sigma_1$}};
		\draw[] (4.6,3.6) node {\textcolor{blue!60}{$\Sigma_2$}};
		\draw[] (2.8,2.55) node {\textcolor{orange}{$\Sigma_{AB}$}};
	\end{tikzpicture}
	\caption{In this figure we show the case when the BPE$(A,B)$ corresponds to the EWCS $\Sigma_{AB}$ anchored on the cutoff brane (or the KR brane in the KR braneworld or AdS/BCFT context). Here the surfaces $\Sigma_1, \Sigma_2$ and $\Sigma_{AB}$ are all minimal surfaces vertically anchored on the cutoff brane and the dashed lines are their geodesic extensions. Solving the balance requirement one can determine the generalized island regions Ir$(A)$, Ir$(B)$, Ir$(A_1)$ and Ir$(B_1)$ which are shown in this figure (see section 5.2.2 in \cite{Basu:2023wmv} for details).
	}
	\label{fig:ABcontribution}
\end{figure}
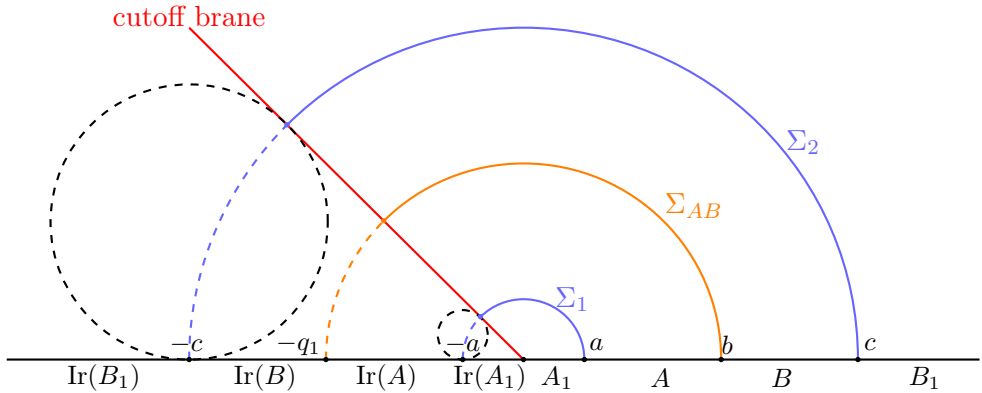

The balance requirement in island phase is still given by\footnote{Note that in \cite{Basu:2023wmv} the balance requirement was also expressed as $\mathcal{I}(A\text{Ir}(A),B\text{Ir}(B)B_1\text{Ir}(B_1))=\mathcal{I}(B\text{Ir}(B),A\text{Ir}(A)A_1\text{Ir}(A_1))$ as the contribution representation and the two-body correlation representation are assumed to be equivalent.}
\begin{align}\label{gbc1}
	\textit{adjacent cases}:
	\quad s_{AA_1}(A)=s_{BB_1}(B)\,.		
\end{align}
Then the contribution $s_{AA_1}(A)$ can be computed by applying the generalized ALC formula \eqref{galc1} to the region $AA_1$, which is
\begin{align}\label{GALC2}
	s_{AA_1}(A)=\frac{1}{2}\left[ \tilde S_{A\text{Ir}(A)A_1\text{Ir}(A_1)}+\tilde S_{A\text{Ir}(A)}- \tilde S_{A_1\text{Ir}(A_1)}\right]\,.
\end{align}
This can further be calculated by the two-point functions according to the basic proposal 2. Similarly we can calculate $s_{BB_1}(B)$ and then solve the balance requirement. The solution determines the partition of $A_1B_1$ and all the generalized island regions $\text{Ir}(A_1)$ and $\text{Ir}(B_1)$ (see Fig.\ref{fig:ABcontribution} and \cite{Basu:2023wmv} for more details). Then the BPE$(A,B)$ is just given by $s_{AA_1}(A)|_{balanced}$. Note that, \eqref{gbc1} may have different solutions under different assignment of the ownerless island regions, which correspond to different saddles of the EWCS, and we should chose the solution that gives the minimal BPE$(A,B)$. In summary, the calculations of the BPE \cite{Basu:2023wmv} in island phase are based on the GALC formula and the basic proposal 1 and 2.

\subsection{Entanglement contribution in island phase revisited}

Nevertheless, according to our PEE threads configuration for island phase, the above derivation \eqref{galc1} for the entanglement contribution $s_{AB}(A)$ has two mistakes:
\begin{itemize}
	\item Firstly in the first line, as we have discussed before the two representations are not equivalent, i.e. 
	\begin{equation}
		s_{AB}(A)
		\neq\mathcal{I}(A\text{Ir}(A),C)\,,
	\end{equation}
	at least for the cases where $AB\text{Ir}(AB)$ is not a single interval, since there are PEE threads connecting points inside $C$ that contribute to $S_{AB}$. 
	
	\item  Secondly, from the second line to the third line, according to the basic proposal 2 the PEE $\mathcal{I}(\gamma,\bar{\gamma})$ does not equal to the multi-point correlation functions of twist operators $\tilde{S}_{\gamma}$ if $\gamma$ is not a single interval, i.e.
	\begin{equation}
		\mathcal{I}(AB\text{Ir}(AB) , C)\neq	\tilde{S}_{AB\text{Ir}(AB)}\,.
	\end{equation}
\end{itemize}
Due to the above mistakes, the derivation \eqref{galc1} for $s_{AB}(A)$ (or $s_{AA_1}(A)$) does not hold in general. 

Interestingly, in all the configurations considered in \cite{Basu:2023wmv} (including the simple case of Fig.\ref{fig:ABcontribution}), $A\text{Ir}(A) A_1 \text{Ir}(A_1)$ is always a single interval. Let us consider the configuration shown in Fig.\ref{fig:ABcontribution} and analyze the entanglement contribution for $AA_1$. In this case $S_{AA_1}$ is given by
\begin{align}\label{paragraph}
	S_{AA_1}=\tilde{S}_{AA_1\text{Ir}A\text{Ir}(A_1)}=\frac{\text{Area}(\Sigma_{AB})}{4G}\,,
\end{align}
which equals to the number of the intersections between $\Sigma_{AB}$ and the PEE threads connecting $AA_1\text{Ir}A\text{Ir}(A_1)$ and its complement. Then it seems natural to take the number of intersections between  $\Sigma_{AB}$ and the PEE threads connecting $A\text{Ir}(A)$ and $B\text{Ir}(B) B_1 \text{Ir}(B_1)$ as the contribution $s_{AA_1}(A)$. Nevertheless, the $s_{AA_1}(A)$ defined in such a way is smaller than the area of the EWCS, i.e. ${\text{Area}(\Sigma_{AB})}/{4G}$, since there are PEE threads emanating from $A_1\text{Ir}(A_1)$ and intersecting with $\Sigma_{AB}$ are not included. In the next subsection we will show that this definition of entanglement contribution is not equivalent to the GALC formula.

Here we would like to stress that, even the steps in the derivation \eqref{galc1} is not correct in general, the generalized ALC formula is still a reasonable proposal for entanglement contribution in island phase. The reason is that, the linear combination in the ALC formula is additive for any physical quantity that can be defined on one-dimensional spatial regions. For example, let us consider an physical quantity $\mathcal{P}_{\mathcal{R}}$ which can be defined on any region $\mathcal{R}$. Consider the region $A=\alpha_L \cup\alpha\cup \alpha_R$ on an infinite line, we can always define the contribution from the sub-region $\alpha$ to $P_A$ as
\begin{equation}\label{ALCP}
	\text{ALC proposal: } p_A\left(\alpha\right)=\frac{1}{2}\left(P_{\alpha_L\cup \alpha}+P_{\alpha\cup \alpha_R}-P_{\alpha_L}-P_{\alpha_R}\right)\,.
\end{equation}
We can divide $\alpha$ to be $\alpha=\alpha_1\cup\alpha_2$ and $A$ is made of the four sub-intervals $\alpha_L \cup\alpha_1 \cup\alpha_2 \cup\alpha_R$ in a row. Then according to the ALC proposal we can compute
\begin{align}
	p_A\left(\alpha_1\right)=\frac{1}{2}\left(P_{\alpha_L\cup \alpha_1}+P_{\alpha\cup \alpha_R}-P_{\alpha_L}-P_{\alpha_2\cup\alpha_R}\right)\,,
	\cr
	p_A\left(\alpha_2\right)=\frac{1}{2}\left(P_{\alpha_L\cup \alpha}+P_{\alpha_2\cup \alpha_R}-P_{\alpha_L\cup\alpha_1}-P_{\alpha_R}\right)\,,
\end{align}
and find that the additivity always hold regardless of the way we define the quantity $P_A$,
\begin{align}
	p_A\left(\alpha\right)=p_A\left(\alpha_1\right)+p_A\left(\alpha_2\right)\,.
\end{align}
One can also check that by construction we always have
\begin{align}
	p_A\left(\alpha_L\right)+p_A\left(\alpha\right)+p_A\left(\alpha_R\right)=P_A\,,
\end{align}
for any $\alpha$. 

Since we want to calculate the entanglement contribution from sub-regions in island phase, and the entanglement entropy in general can be expressed as $S_{AA_1}=\tilde{S}_{AA_1\is (AA_1)}$ (here $\text{Ir}(AA_1)=\is (AA_1)$), it is a good choice to consider this quantity $P_\alpha=\tilde{S}_{\alpha\text{Ir}{\alpha}}$ for any sub-region $\alpha$ of $AA_1$. Then when taking $\alpha=A$, \eqref{ALCP} is exactly the GALC formula \eqref{GALC2} and the entanglement contributions are given by
\begin{align}
	s_{AA_1}(A) =& \frac{1}{2}\left[ \tilde S_{A\text{Ir}(A)A_1\text{Ir}(A_1)}+\tilde S_{A\text{Ir}(A)}- \tilde S_{A_1\text{Ir}(A_1)}\right] 
	\cr
	= & \frac{1}{2}\left[ \frac{\text{Area}(\Sigma_{AB})}{4G}+\left(\frac{\text{Area}(\Sigma_{AB})}{4G}+\frac{\text{Area}(\Sigma_{1})}{4G}\right)- \frac{\text{Area}(\Sigma_{1})}{4G}\right]
	\cr
	=& \frac{\text{Area}(\Sigma_{AB})}{4G}\,,
	\\
	s_{AA_1}(A_1) =& \frac{1}{2}\left[ \tilde S_{A\text{Ir}(A)A_1\text{Ir}(A_1)}+\tilde S_{A_1\text{Ir}(A_1)}- \tilde S_{A\text{Ir}(A)}\right] 
	\cr
	= & \frac{1}{2}\left[ \frac{\text{Area}(\Sigma_{AB})}{4G}+\frac{\text{Area}(\Sigma_{1})}{4G}-\left(\frac{\text{Area}(\Sigma_{AB})}{4G}+\frac{\text{Area}(\Sigma_{1})}{4G}\right)\right]
	\cr
	=& 0\,.
\end{align}
As we can see, the contribution from $A_1\text{Ir}(A_1)$ to $S_{AA_1}$ is zero, and the area of the EWCS is reproduced.

For the special case that $AA_1\is (AA_1)$ is a single interval, according to the basic proposal 1, the entanglement entropy is also captured by
\begin{align}
	S_{AA_1}=\mathcal{I}(AA_1\is (AA_1), C)\,, \qquad C=BB_1\is (BB_1)\,.
\end{align}
In this case, there exists another natural choice $P_{\alpha}=\mathcal{I}(\alpha\text{Ir}(\alpha),C)$, which gives the following entanglement contribution
\begin{align}
	s_{AA_1}(A)=&\frac{1}{2}\left(\mathcal{I}(A\text{Ir}(A),C)+\mathcal{I}(AA_1\is (AA_1),C)-\mathcal{I}(A_1\text{Ir}(A_1),C)\right)
	\cr
	=&\mathcal{I}(A\text{Ir}(A),C)\,.
\end{align}
This definition seems more consistent with our PEE threads configuration, nevertheless it does not apply to generic cases and the entanglement contribution $s_{AA_1}(A)$ under balanced condition does not reproduce the EWCS since
\begin{equation}
	\mathcal{I}(A\text{Ir}(A),C)<\mathcal{I}(AA_1\text{Ir}(A)\text{Ir}(A_1),C)=\frac{\text{Area}(\Sigma_{AB})}{4G}\,.
\end{equation}
A thorough analysis for the EWCS in island phase using the bulk PEE threads configuration is needed to understand the failure of this proposal for entanglement contribution, we leave this topic for future investigation.

\subsection{Non-vanishing PEE and the independence between the two basic proposals}
The above mentioned mistakes also lead to another wrong claim. In \cite{Basu:2023wmv}, the basic proposal 2 \eqref{Basic proposal 2} was expressed as:
\begin{equation}\label{StildeA}
	\mathcal{I}(A\text{Ir}(A),EF)=\tilde{S}_{[d,b]}+\tilde{S}_{[c,a]}\,.
\end{equation}
based on the wrong assumption $\tilde{S}_{A\text{Ir}(A)}=\mathcal{I}(A\text{Ir}(A),EF)$. According to the additive property of the PEE and the basic proposal 1, we have,
\begin{align}
	\mathcal{I}(A\text{Ir}(A),EF)=&\mathcal{I}(A\text{Ir}(A),E)+\mathcal{I}(A\text{Ir}(A),F)\,,
	\cr
	\tilde{S}_{[d,b]}+\tilde{S}_{[c,a]}=&\mathcal{I}(A\text{Ir}(A)F,E)+\mathcal{I}(A\text{Ir}(A)E,F)
	\cr
	=&\mathcal{I}(A\text{Ir}(A),E)+\mathcal{I}(A\text{Ir}(A),F)+2\mathcal{I}(E,F)\,.
\end{align}
Then \eqref{StildeA} means $\mathcal{I}(E,F)=0$, and it was used to demonstrate that the basic proposal 2 is a result of the basic proposal one (see (46) and (47) in \cite{Basu:2023wmv}). Nevertheless, this claim is wrong and inconsistent with our previous result \eqref{regulatedPEE}, which is that $\mathcal{I}(E,F)= \mathcal{I}^{\text{CFT}}(E,F)\neq 0$. So the two basic proposals are indeed independent from each other as we have discussed in section \ref{sec.PEEinislandphase}.

\section{Summary and discussions}
In this paper we have discussed the entanglement structure of the island phase using the PEE threads in the context of holographic Weyl transformed CFT$_2$. The essential reason for the emergence of entanglement islands in this model is that, the dynamics of the Weyl scalar is gravitational hence the Weyl transformed region is coupled to gravity, and the Weyl transformation introduces finite cutoff, hence there exist configurations with entanglement islands in the gravitational region which give smaller entanglement entropy according to the island formula \eqref{island formula 1}. Also the finite cutoff indicates that physics at the scale smaller than the cutoff becomes meaningless, hence smears the local description of boundary points. Compared with the PEE threads configuration in Poincar\'e AdS$_3$, the new PEE threads configuration we proposed in the island phase : 1) replaces the boundary points with cutoff spheres whose radius are determined by the finite cutoff scale, 2) requires the homologous surface to anchor on the cutoff spheres instead of the boundary points and 3) leaves the distribution of the bulk PEE threads unchanged.

In the vacuum state of holographic CFT$_2$ that duals to AdS$_3$, the two-body PEE $\mathcal{I}(A,B)$ has a very simple interpretation, which is just the number of PEE threads connecting $A$ and $B$. Nevertheless, the PEE in island phase was not well understood. Under our new setup of PEE thread configurations, we can explore $\mathcal{I}(A,B)$ in island phases, and the results are summarized in \eqref{regulatedPEE}. Since the endpoints of the boundary intervals are replaced by the corresponding cutoff spheres, it is natural to interpret the PEE between a single interval $A$ and its complement $\bar{A}$ as the number of PEE threads that pass through the bottle neck, i.e. the minimal homologous surface anchored on the two cutoff spheres at $\partial A$, to connect $A$ and $\bar{A}$. This exactly matches the basic proposal 1.  Based on the PEE $\mathcal{I}(A,\bar{A})$ for any single interval $A$ and the additivity, we calculated the PEE between any two intervals $A$ and $B$, which can be adjacent or disconnected. Note that when $A$ and $B$ are disconnected, $\mathcal{I}(A,B)$ is the same as the $\mathcal{I}^{\text{CFT}}(A,B)$ in the holographic CFT$_2$, which confirms our setup that the Weyl transformation only removes small scale entanglements below the cutoff scale, while leave the large scale entanglement structure unchanged.

It is also a natural extension to take the holographic four-point (or higher-point) function of twist operators as the area of the minimal homologous surface anchored on the four corresponding cutoff spheres. Similarly to the RT surface for two disconnected intervals in holographic CFT$_2$, there could be a phase transition between phases with connected and disconnected entanglement wedge. Applying this generalization to the four twist operators at the boundary points of $A\cup \text{Ir}(A)$, which is in the phase with connected entanglement wedge, we get the basic proposal 2. Furthermore, by allowing the homologous surface to anchor on any cutoff sphere other than those centered at the boundary points of the region under consideration, we reproduced the island formula by minimizing the number of intersections between the homologous surface and PEE threads.

Combining the Proposal 2 and our PEE threads configuration, we conclude that: 1) the normalization property $\tilde{S}_{A\text{Ir}(A)}=\mathcal{I}(A\text{Ir}(A),C)$ does not hold in general; 2) counting the PEE threads emanating from a sub-region of $A$ to $\bar{A}$ is not enough to capture its contribution to the entanglement entropy $S_{A}$. These conclusions give us new understanding on the three conditions, i.e. the generalized ALC formula and the two basic proposals, based on which the BPEs in island phase were computed in \cite{Basu:2023wmv}. 
We also clarify some incorrect statements about the derivation of the GALC formula and the independence between the two basic proposals in \cite{Basu:2023wmv}. 

Although we have given some arguments in favor of the GALC formula to define the entanglement contribution, the role of the GALC formula and the balance requirements in reproducing the EWCS is still not fully understood. An interesting future direction is to give a thorough analysis for the EWCS by optimizing the number of intersections with the PEE threads, which may give an interpretation for the GALC formula and the balance requirements in terms of an optimization problem.

\acknowledgments

We thank Yizhou Lu, Jiong Lin for helpful discussions. The authors thank the Shing-Tung Yau Center of Southeast University for support. HZ was supported by SEU Innovation Capability Enhancement Plan for Doctoral Students (Grant No.CXJH\_SEU 24137).


	\bibliographystyle{JHEP}
	\bibliography{lmbib}
	
\end{document}